\documentclass[useAMS,usenatbib]{mn2e}

\usepackage{aas_macros,epsfig}

\title[Faint $z\approx 6$ Ly-$\alpha$ Line Emitters in the  HUDF]
{The GLARE Survey II. Faint $z\approx6$ Ly-$\alpha$ Line Emitters in the HUDF}
\author[E R Stanway et al.]{Elizabeth R.~Stanway$^{1}$\thanks{Current Address: H H Wills Physics Laboratory, Bristol, BS8 1TL, UK}, Andrew~J.~Bunker$^{2,3}$, Karl Glazebrook$^{4}$, \newauthor 
Roberto G.~Abraham$^{5}$, James Rhoads$^{6}$, Sangeeta Malhotra$^{6}$, \newauthor
David Crampton$^{7}$, Matthew Colless$^8$, Kuenley Chiu,$^{3,4}$\\
$^1$ Astronomy Department, University of Wisconsin-Madison, Sterling Hall, Madison, WI, 53726, USA\\
$^2$ Institute of Astronomy, Madingley Road, Cambridge, CB3\,OHA, UK\\
$^3$ School of Physics, University of Exeter, Stocker Road, Exeter, EX4\,4QL, UK\\
$^4$ Department of Physics and Astronomy, John Hopkins University, 3400 N Charles St, Baltimore, MD 21218, USA\\
$^5$ Dept of Astronomy \& Astrophysics, University of Toronto, 60 St. George St, Toronto, ON, M5S 3H8, Canada\\
$^6$ Arizona State University, Department of Physics \& Astronomy, Box 871504, Temp, AZ\,85287, USA\\ 
$^7$ Dominion Astrophysical Observatory, 5071 W Saanich Rd, Victoria, V9E 2E7, Canada\\
$^8$ Anglo-Australian Observatory, P.O.Box 296, Epping, NSW 1710,  Australia}

\begin{document}

\date{Accepted . Received ; in original form }

\pagerange{\pageref{firstpage}--\pageref{lastpage}} \pubyear{}

\maketitle

\label{firstpage}

\begin{abstract}
  The galaxy population at $z\approx6$ has been the subject of intense
  study in recent years, culminating in the Hubble Ultra Deep Field
  (HUDF) -- the deepest imaging survey yet.  A large number of high
  redshift galaxy candidates have been identified within the HUDF, but
  until now analysis of their properties has been hampered by the
  difficulty of obtaining spectroscopic redshifts for these faint
  galaxies.  Our ``Gemini Lyman-Alpha at Reionisation Era'' (GLARE)
  project has been designed to undertake spectroscopic follow up of
  faint ($z'<28.5$) $i'$-drop galaxies at $z\approx 6$ in the HUDF.
  In a previous paper we presented preliminary results from the first
  7.5 hours of data from GLARE.  In this paper we detail the complete
  survey. We have now obtained 36 hours of spectroscopy on a single
  GMOS slitmask from Gemini-South, with a spectral resolution of
  $\lambda/\Delta\lambda_{\rm FWHM} \approx 1000$. We identify five
  strong Lyman-$\alpha$ emitters at $z>5.5$, and a further nine
  possible line emitters with detections at lower significance.  We
  also place tight constraints on the equivalent width of
  Lyman-$\alpha$ emission for a further ten $i'$-drop galaxies and
  examine the equivalent width distribution of this faint
  spectroscopic sample of $z\approx6$ galaxies. We find that the
  fraction of galaxies with little or no emission is similar to that
  at $z\approx3$, but that the $z\approx6$ population has a tail of
  sources with high rest frame equivalent widths. Possible
  explanations for this effect include a tendency towards stronger
  line emission in faint sources, which may arise from extreme youth
  or low metallicity in the Lyman-break population at high redshift,
  or possibly a top-heavy initial mass function.

\end{abstract}


\section{Introduction}
\label{sec:intro}

The Hubble Ultra Deep Field \citep[HUDF,][]{astroph0607632} opened a new window on the early
universe. The exceptionally deep, multiwavelength data provided
opportunities to study colour-selected samples of high redshift galaxy
candidates with modest luminosities more typical of the general galaxy
population. Previously, only the most luminous `tip of the iceberg'
had been accessible.  By pushing to redshifts around 6 with the
$z'$-band filter and the $i'$-drop Lyman break selection technique \citep[e.g.][]{2004MNRAS.355..374B}, the HUDF
explored the end of the reionization epoch signaled by the
Gunn-Peterson HI absorption trough in QSOs \citep{2001AJ....122.2850B}.

Nonetheless, the use of the Lyman break colour selection criterion to
isolate star-forming galaxies at a specific redshift, first developed
to study $z\approx3$ galaxies \citep{1995AJ....110.2519S} and extended
to $z\approx6$ candidates in analysis of the Great Observatories
Origins Deep Survey (GOODS) and HUDF fields
\citep[e.g.][]{2003MNRAS.342..439S,2006ApJ...653..53B}, presents
challenges. Before the properties of the population can be
meaningfully discussed, the selection function itself must be
understood.  Estimates must be obtained of the contaminant fraction
and the redshift distribution of $i'$-drop galaxies.

Many colour-selected galaxies are significantly fainter than the
conventional spectroscopic limit of today's large telescopes. The HUDF
reaches limits of $z'_{AB}=28.5$ (10\,$\sigma$) -- a depth that, until
{\em JWST} becomes available, requires unreasonable exposure times to
obtain a high signal-to-noise ratio (S/N) spectrum for the continuum.
However, line emission (e.g. Lyman-$\alpha$) may be within reach of
ultra-deep moderate-dispersion spectroscopy on 8-10\,m telescopes even
for the faintest galaxies, provided the equivalent width of the line
is large enough.
%Nonetheless, spectroscopy of these sources can yield valuable results.
If the properties of the star-forming population at $z\approx6$ are
similar to those of the Lyman Break Galaxy (LBG) population at
$z\approx3$ \citep{2003ApJ...592..728S}, then half the sources are
expected to show the resonant Lyman-$\alpha$ transition in emission.
Detection of a Lyman-$\alpha$ emission line fixes the redshift of a
source, while detection or constraints on nearby high ionisation
emission lines can quantify contribution by an AGN. Furthermore,
spectroscopy can enable the identification of lower-redshift
contaminants in the sample which will not emit Lyman-$\alpha$ photons
at these wavelengths,
but may show other emission lines. When a population of Lyman-$\alpha$
emitters is studied, the distribution of equivalent widths is
sensitive to the stellar initial mass function (since the
Lyman-$\alpha$ transition is excited by emission from hot, massive,
short-lived stars), and also to the fraction of neutral gas in the
intergalactic medium (IGM) and the emergent photon fraction
\citep{2004ApJ...617L...5M}.

\begin{figure}
\begin{center}
\resizebox{0.75\columnwidth}{!}{\includegraphics{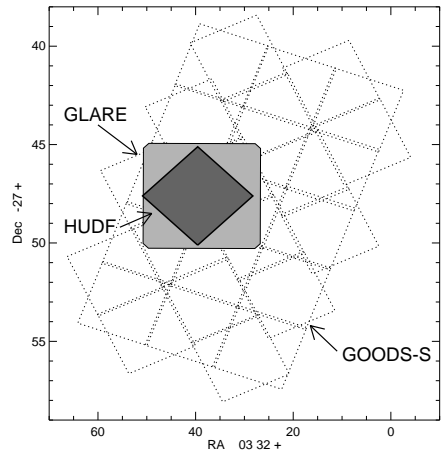}}
\end{center}
\caption{The geometry of the UDF and GOODS-S fields (taken at two position angles), and the layout of
  the GLARE mask. GLARE encompasses the entire UDF field, and some
  parts of the GOODS-S field}\label{fig:layout}
\end{figure}

In this paper we present results from the Gemini Lyman-Alpha at
Reionisation Era (GLARE) project. This program used the 8-m Gemini
South telescope to obtain extremely deep spectroscopy on a single
slitmask, centered on the HUDF (Figure \ref{fig:layout}). By obtaining
extremely long exposures using a telescope with a large collecting
area, we aimed to study continuum-selected galaxies fainter than those
targeted by any other survey, and to quantify the line emission
properties of our $i'$-drop sample in the HUDF. We presented initial
results from this programme in \citet[][ hereafter Paper
I]{2004ApJ...604L..13S}, which was based on 7.5\,hours of on-source
exposure taken at low spectral resolution
($\lambda/\Delta\lambda\approx 500$). In this paper we present an
analysis of the twenty-four $i'$-drop selected $5.6 < z < 7.0$
candidates targeted for 36\,hours of spectroscopy at higher resolving
power ($\lambda/\Delta\lambda\approx 1200$). In section
\ref{sec:glare} we describe the GLARE program, and in section
\ref{sec:emitters} we present the results of our observing campaign.
In section \ref{sec:EW} we analyse the equivalent width distribution
of the GLARE line emitters, and in section \ref{sec:disc} we discuss
the implications of this distribution in the context of other work in
this field. Finally in section \ref{sec:conc} we present our
conclusions.

%\textit{The progress in redshift 6 work. The spectroscopy of line emitters. Isolated spectroscopy of $i'$-drop galaxies. The difficulty in obtaining redshifts for faint sources. The need to quantify Lyman-alpha contamination, to identify AGN, to verify the redshift distribution of $i'$-drop sources.}
%
%\textit{Discuss HUDF, Define GLARE, mention GRAPES}

We adopt 
the following cosmology: a flat Universe 
with $\Omega_{\Lambda}=0.7$, $\Omega_{M}=0.3$ and $H_{0}=70 h_{70} 
{\rm km\,s}^{-1}\,{\rm Mpc}^{-1}$. All magnitudes are 
quoted in the AB system (Oke \& Gunn 1983).

\section{The GLARE Project}
\label{sec:glare}

  \subsection{Candidate Selection}
  \label{sec:cands}

%  \def\th{\hbox{$^{\rm th}$}} The spectra described in this paper were
%  obtained as part of the GLARE project -- an effort to obtain
 % extremely deep spectra of a single slitmask, targeting faint
 % $z\approx6$ candidate galaxies in the Hubble Ultra Deep Field with
 % the Gemini Multi-Object Spectrograph (GMOS, Hook et al.\ 2003) on
 % Gemini-South.
  
  The objectives of the GLARE project were to determine the redshift
  and line emission properties of our targets, reaching low equivalent
  width limits on Lyman-$\alpha$ emission for a uniformly-selected
  sample of $i'$-drop Lyman-break galaxies
  \citep{2003MNRAS.342..439S,2004ApJ...606L..25B,2004MNRAS.355..374B}.
  The $i'$-drop colour selection is sensitive to galaxies at
  $5.6<z<7.0$, although the selection sensitivity falls off rapidly
  above $z\approx6.5$ (see Stanway et al.\ 2003).  With this data we
  aim to probe galaxies at the end of the epoch of reionisation, and
  compare their properties to comparable galaxy samples at lower
  redshift, in order to explore possible evolution in the nature of
  the star-forming galaxy population.
  
  Initial results obtained from this survey were reported in
  \citet{2004ApJ...604L..13S}, which presented 7.5\,hour spectra of
  three Lyman-$\alpha$ line emitters observed on the 2003 GLARE
  slitmask.  Selection of targets for our 2003 observations was made
  possible by the early release of a list of red sources in the HUDF,
  based on observations to one third of its final depth, and
  supplemented by $i'$-drop galaxies selected from the shallower, but
  wider field, GOODS survey of the same region (Stanway 2004).

  Between observations of our 2003 GLARE mask, and the start of
  semester 2004B, the full HUDF data were made public and the GLARE
  slitmask was redesigned accordingly. Sources with
  $(i'-z')_{AB}\ge1.3$ and $z'_{AB}\le28.5$ in the HUDF imaging
  \citep[i.e. the catalog of ][]{2004MNRAS.355..374B} were assigned the
  highest priority for spectroscopic follow-up
  and fifteen such sources were placed on the slitmask. The agreement
  is excellent between the HUDF $i'$-drop discovery catalog of
  \citet[][]{2004MNRAS.355..374B}, which we use as our GLARE source
  list, and the recent data paper by the HUDF team
  \citep{astroph0607632}. All the robust $i'$-drops targetted on the
  GLARE slitmask were are also identified by
  \citeauthor{astroph0607632}

  In order to maximise use of the mask, additional objects were targetted.
 Seven slits were placed on candidates (including the two emitters
  identified in Paper~I, GLARE 3001 \& 3011) from a brighter selection
  with $(i'-z')_{AB}\ge1.3$ and $z'_{AB}\le27.5$, selected from the
  GOODS imaging (Stanway 2004), primarily at the edges of the slitmask
  and lying outside the HUDF region. A further three slits were placed
  on candidates with colours lying marginally outside of our GOODS
  selection criteria, either in colour (i.e.\ $1.0<i'-z'<1.3$) or
  magnitude (i.e. $z'>27.5$), or in noisy regions of the GOODS images,
  and two slits on sources from the original UDF early release list of
  red sources that did not meet our final criteria. Two slits were
  placed on $z\approx5$ $v$-band dropout candidates, and two on known
  low redshift [OII]\,3727\,\AA\ emitters, previously observed by
  VLT/FORS2 \citep{2006astro.ph..1367V} and used as a check on flux
  calibration.  Finally, five slits were placed on alignment stars to
  ensure accurate positioning of the mask, and six slits were used to
  conduct a blank sky survey for serendipitous sources.  The
  composition of the 2004 GLARE slitmask is summarised in table
  \ref{tab:mask}.

\begin{table}
\begin{tabular}{ccl}
ID & No of Slits & Description \\
\hline\hline
%1000-1999 & 15 & HUDF, $i'-z'\ge1.3$, $z'\le28.5$  \\
%          &  2 & HUDF, other red sources           \\
%2000-2999 &  5 & Alignment Stars                   \\
%3000-3999 &  7 & GOODS $i'-z'\ge1.3$, $z'\le27.5$  \\
%4000-4999 &  2 & GOODS $v-i'$ selected, $z=5$ cands\\
%5000-5999 &  2 & Known, low z, [OII] emitters      \\
%6000-6999 &  6 & Blanks sky slits                  \\
%8000-8999 &  3 & GOODS marginal $i'$-drop candidates\\
1{\sc XXX} & 15 & HUDF, $i'-z'\ge1.3$, $z'\le28.5$  \\
     &  2 & HUDF, other red sources           \\
2{\sc XXX} &  5 & Alignment Stars                   \\
3{\sc XXX} &  7 & GOODS $i'-z'\ge1.3$, $z'\le27.5$  \\
4{\sc XXX} &  2 & GOODS $v-i'$ selected, $z=5$ cands\\
5{\sc XXX} &  2 & Known, low z, [OII] emitters      \\
6{\sc XXX} &  6 & Blanks sky slits                  \\
7{\sc XXX} &  3 & GOODS marginal $i'$-drop candidates\\
\hline
& 42 & Total Number of Slits on Mask     \\   
\end{tabular}
\caption{The composition of the 2004 GLARE slitmask. 1XXX indicates a GLARE identifier number in the range 1000-1999, etc.}\label{tab:mask}
\end{table}

%\textit{mention November 2003 stack of first mask? The loss of one channel in some of the data?}

  Both the $z'_{AB}=27.5$ cut applied in the GOODS data and the
  $z'_{AB}=28.5$ cut applied in the HUDF correspond to a signal to
  noise of approximately 8. We chose to work well above the detection
  limit in order to have confidence in the reality and nature of our
  candidate sources (many of which are detected only in this one
  band). In the event of non-detection in the $i'$-band we use the
  limiting magnitudes $i'_{AB}=29.15$ (GOODS) and $i'_{AB}=30.4$
  (HUDF) corresponding to 3\,$\sigma$ variation in the sky background,
  as measured on the images. All sources were required to be
  undetected in the available $b$ (F435W) imaging but faint detections
  in the deep $v$ (F606W) filter (which lies above the Lyman limit at
  $z\approx6$, and which is present in several spectroscopically
  confirmed $z>5.6$ galaxies) were permitted. Completely unresolved
  sources were omitted from the selection although at faint
  magnitudes, the dividing line between unresolved and
  slightly-resolved sources becomes blurred.
  
  Of the line emitter candidates presented in section
  \ref{sec:emitters}, only GLARE 1042 and GLARE 1040 lie within the
  NICMOS HUDF field \citep{2005AJ....130....1T}. Both have
  near-infrared colours consistent with a high redshift
  interpretation, as indeed did all the $i'$-drop sources in this
  field discussed by \citet{2005MNRAS.359.1184S}.

  \subsection{Observations and Data Analysis}
  \label{sec:data}
  
  The 2004 GLARE slitmask was observed in semester 2004B using the
  GMOS spectrograph on Gemini-South (Hook et al.\ 2003). This mask was
  observed at higher spectral resolution than the 2003 GLARE mask,
  using the GMOS-S R600 grating rather than the R150, blazed at an
  angle of 48.5$\degr$ giving a central wavelength of $\sim8700$\AA .
  As the CCD over-sampled the typical seeing and spectral resolution,
  the image was binned at $2\times2$ pixels so as to reduce the
  readout noise and improve the S/N. After this binning, the spectral
  scale was 0.94\,\AA/pixel, and 0.146\,arcsec/pixel spatially on the
  detector.  The slit width was 1.0\,arcsec, which produced a spectral
  resolution of 6.5\,\AA\ FWHM for objects which fill the slit. Both
  mask and slits were oriented due north. The higher spectral
  resolution of the 2004 GLARE mask ($\lambda/\Delta\lambda\approx
  1200$ compared with $500$ for 2003 GLARE) decreases the fraction of
  the wavelength range adversely affected by OH sky lines, and also
  enables us to better study the profiles of emission lines from the
  targets.  The OG515 filter was used to suppress second order light
  from shorter wavelengths. In order to fill CCD chip gaps and ensure
  full wavelength coverage in the range $\lambda\approx7000-10000$\AA\ 
  three different central wavelength settings were observed (8580\AA,
  8700\AA\ and 8820\AA). The shortest wavelength reached was 6420\AA, while the longest wavelenth surveyed was 10950\AA. 
% and prevent oversampling of resolution elements.
  
  Wavelengths were calibrated from the night sky lines in each slit,
  leading to a solution with an {\em rms} error of approximately
  0.3\AA. Fluxes were calibrated from the broadband magnitudes of the
  alignment stars on the mask, and checked against both line emission
  of known [OII] emitters also observed in VLT/FORS2 spectroscopy that
  had been placed on our mask for verification purposes, and existing
  spectroscopy for the known $z=5.83$ Lyman-$\alpha$ emitter GLARE
  1042
  \citep{2004ApJ...604L..13S,2004ApJ...607..704S,2004ApJ...600L..99D}.
  We estimate a 20\% error on the flux calibration, associated
  with slit losses and centroiding uncertainty in the narrow slits.
%on these faint fluxes, due to a combination
%  of residual sky background noise and Poisson noise on the line emission.

  To optimise the subtraction of night sky emission lines (which
  occupy a large fraction of the spectrum at $>8000$\AA) we used the
  instrument in `nod \& shuffle' mode \citep{gla01,2004astro.ph..2436A}. Each
  exposure was 30 minutes long, nodding every 60 seconds. Hence we are
  able to suppress sky emission that varies on timescales longer than
  one minute. The total exposure time on this mask was 36 hours.
  
  The reduction of nod \& shuffle data involves the subtraction of
  positive and negative spectra, observed in alternate 60 second
  exposures and offset spatially by the telescope nodding. We employed
  slits 2.47 arcseconds in length, nodding by 1.25 arcseconds between
  sub-exposures. Our queue scheduled observations were constrained by
  the requirement that the seeing was less than 0.5 arcseconds FWHM,
  and the nod distance set such that the signals were separated by at
  least twice the seeing disk.  Hence the characteristic signal of an
  emission line comprises a `dipole' signal of positive and negative
  emission at the same wavelength, spatially offset by 1.25
  arcseconds. In visually identifying candidate line emitters, we
  looked for this dipole signature with positive and negative
  channels of comparable strengths; this requirement effectively
  increases the sensitivity of the spectroscopy beyond the formal
  limit, since random background fluctuations are unlikely to produce
  simultaneous signals in both positive and negative channels. 
 
  Dipole signals lying under strong emission sky lines are treated
  with caution; as well as having more Poisson noise, sky line
  variability and charge diffusion at the red end of the spectrum may
  lead to a spurious signal.

  A number of charge traps and CCD artifacts were also masked when the
  individual exposures were combined. These charge traps affect some
  regions of the CCD more severely than others, and so the noise
  properties vary from slitlet to slitlet, and
  also with wavelength. However, the 1 sigma {\em rms} pixel-to-pixel
  variation in the background at 8500\AA\ was
  $1.4\times10^{-19}$\,erg\,cm$^{-2}$\,s$^{-1}$\,\AA$^{-1}$ for the
  $2\times 2$ binned pixels, measured between skylines on the nod \&
  shuffle background-subtracted spectra.  Hence, our sensitivity to an
  emission line extending over 500\,km\,s$^{-1}\times 1$\,arcsec is
  $2.5\times 10^{-18}$\,erg\,cm$^{-2}$\,s$^{-1}$ ($3\,\sigma$ combining
  both nod positions, or $2\,\sigma$ per nod channel).

% (in which
%  the dark current and sky continuum cancel, leaving only Poisson
%  noise on the sky background, and those fluctuations which occur on a
%  timescale shorter than 60\,s).

\section{Line Emitters in the 2004 GLARE Mask}
\label{sec:emitters}

In the 36 hour exposure of the 2004 GLARE mask, we identify five
strong Lyman-$\alpha$ emission line sources \citep[including the three candidates
tentatively proposed in][]{2004ApJ...604L..13S}. We identify a further
four sources which have lower significance detections, but which are
considered possibly
to be real emission lines due to their dipole natures and separation
from possible sky line residuals. Finally we identify five sources with
tentative emission line detections that are considered unlikely to be
real.

  \subsection{Strong Emission Lines}
  \label{sec:good}
  
\begin{figure*}
\begin{center}
\begin{tabular}{ccc}
\resizebox{0.55\columnwidth}{!}{\includegraphics{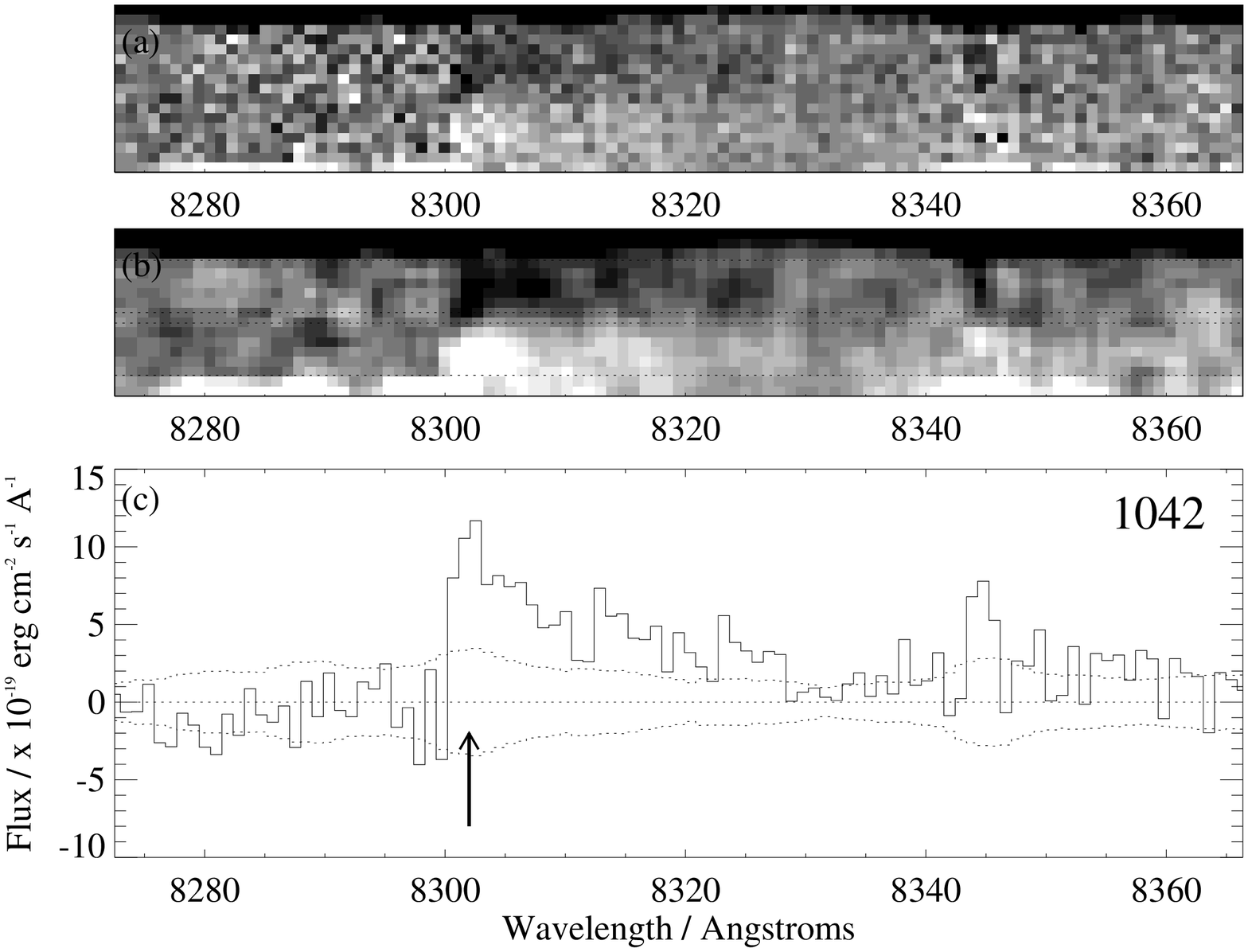}} &
\resizebox{0.55\columnwidth}{!}{\includegraphics{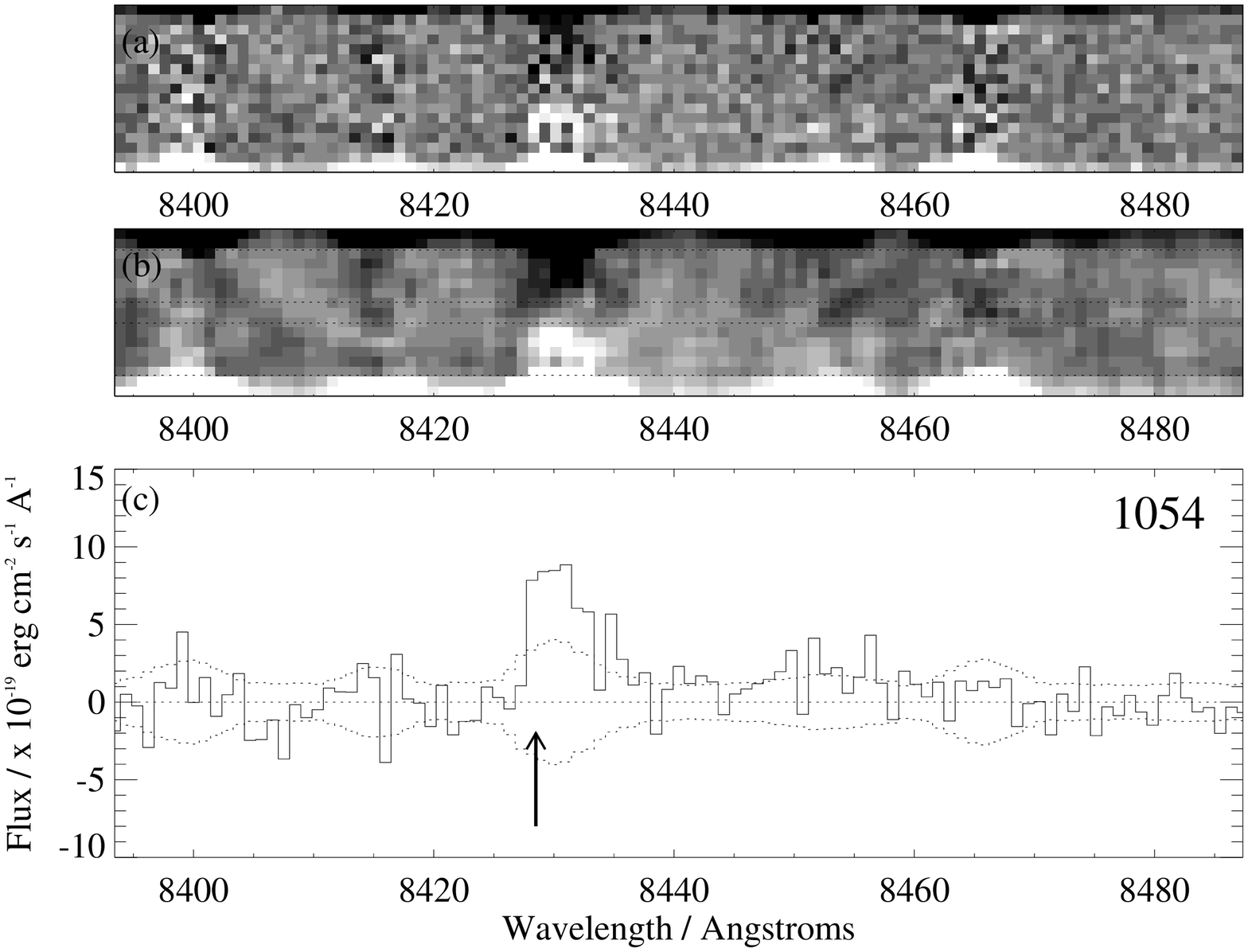}} &
\resizebox{0.55\columnwidth}{!}{\includegraphics{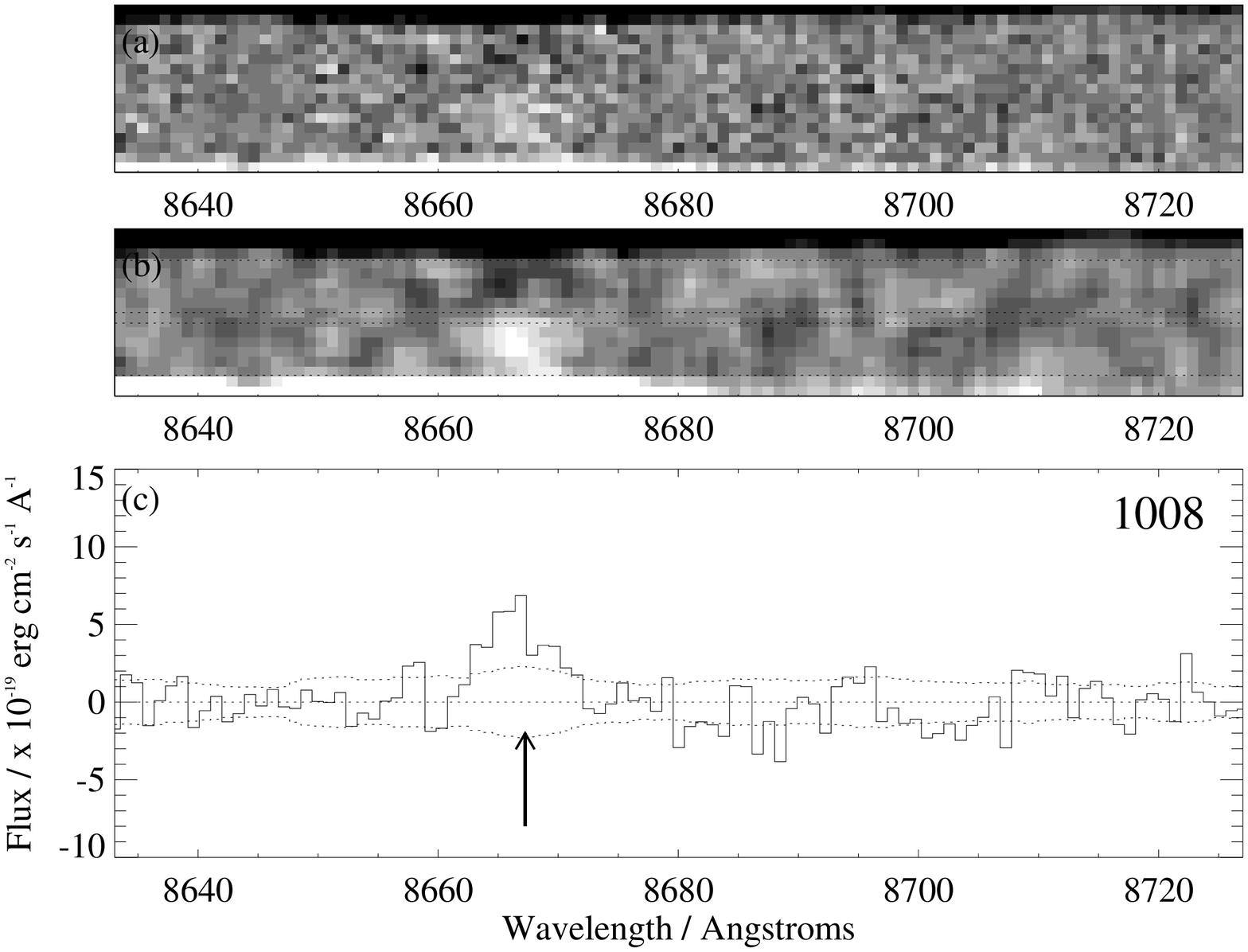}} \\
\resizebox{0.55\columnwidth}{!}{\includegraphics{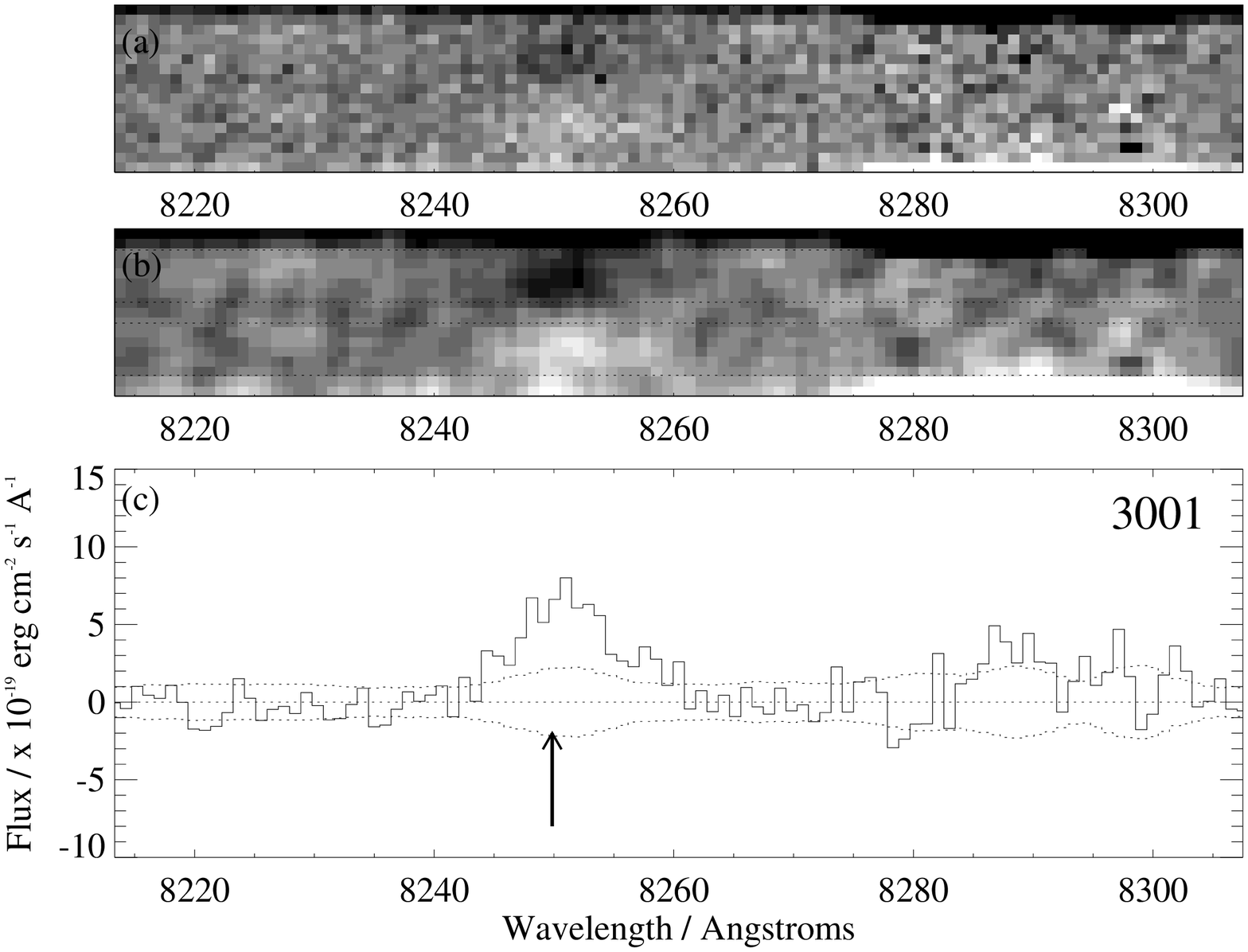}} &
\resizebox{0.55\columnwidth}{!}{\includegraphics{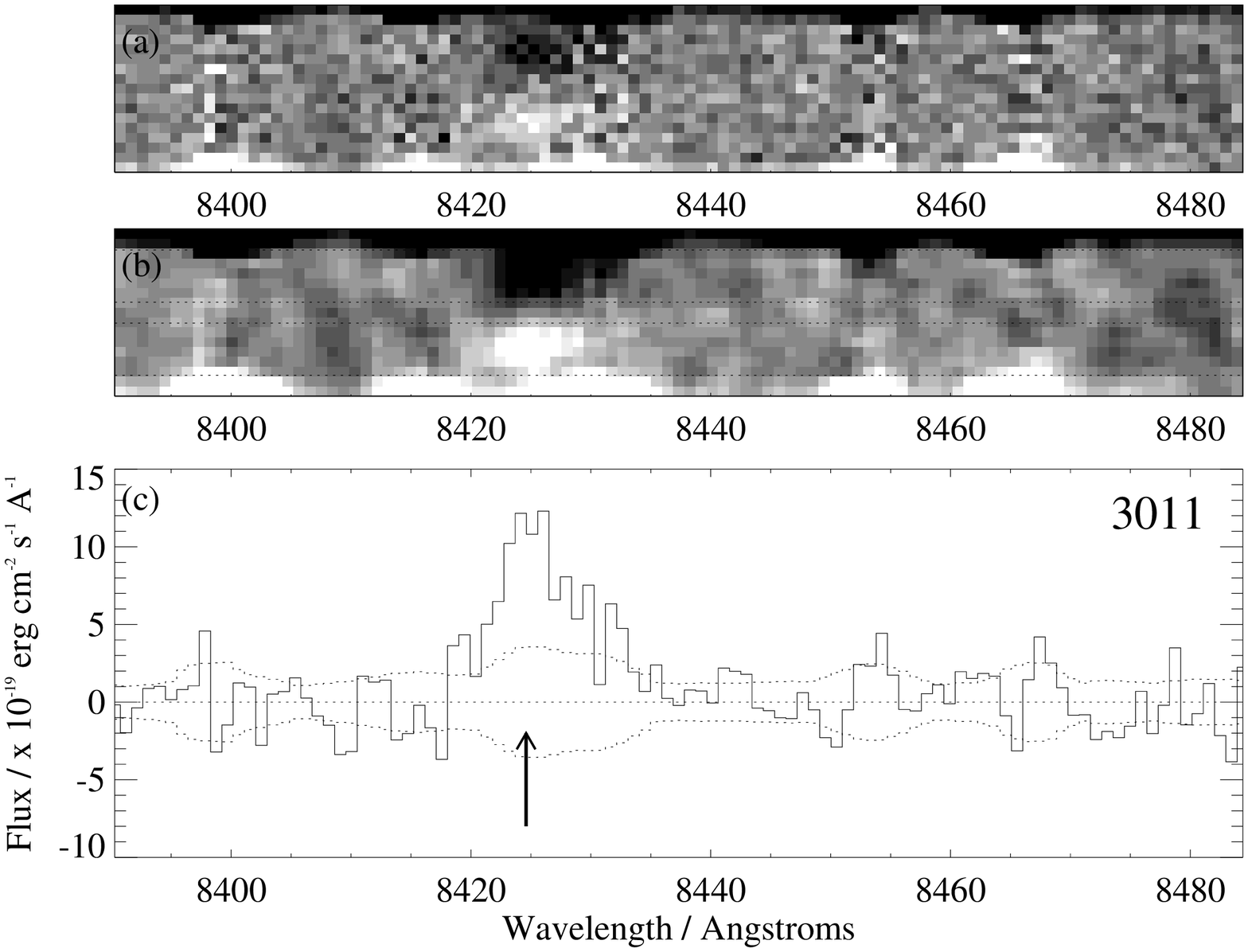}} & \\
\end{tabular}
\end{center}

\caption{One and two dimensional spectra of the strong emission line
  candidates. The upper panel (a) shows the two dimensional spectrum,
  the central panel (b) a spectrum smoothed over three pixels, and the
  lower panel (c) the one dimensional combined spectrum extracted from
  both positive and negative channels. The source ID is shown on the
  right of panel (c) in each case. The 1 sigma standard deviation in
  the background, smoothed over 15\AA\ and including the Poisson noise
  on the emission line flux is indicated with a dotted line. The arrow
  indicates the wavelength of the putative emission
  line.}\label{fig:strong}

\end{figure*}

  Table \ref{tab:strong} lists the properties of the five $z\approx6$
  sources with strong Lyman-$\alpha$ line emission. Figure
  \ref{fig:strong} presents the two dimensional spectra obtained for
  these sources, and the summed flux from positive and negative
  spectral channels. We also placed two known lower-redshift galaxies
  on the 2004 GLARE mask, placing [OII] within our spectral range, as
  a check on our sensitivity and calibration.  These are galaxies GDS
  J033235.79-274734.7 ($z=1.223$) and GDS J033229.63-274511.3
  ($z=1.033$) from the ESO VLT/FORS2 survey of \citet{2005A&A...434...53V}.
  We detect the [OII] emission at
  $\lambda=8285$\,\AA\,\&\,$7577$\,\AA , with line fluxes
  $1.53$\,\&\,$1.52\,\times 10^{-17}\,{\rm erg\,cm^{-2}\,s^{-1}}$,
  respectively.

  For the Lyman-$\alpha$ detections, GLARE sources 1042 and 3001 were
  previously identified as line emitters in Paper I from the 2003
  GLARE mask, \citet{2004ApJ...604L..13S}. GLARE 3011 was tentatively
  identified as a line emitter, and this identification is now
  strongly confirmed. GLARE 1054 and 1008 are presented for the first
  time here.
  
  Since its discovery \citep[SBM03\#1 in ][]{2003MNRAS.342..439S}
  GLARE 1042 at $z=5.83$ has been spectroscopically confirmed
  \citep[][--
  SiD002]{2004ApJ...607..704S,2004ApJ...604L..13S,2004ApJ...600L..99D}
  and extensively studied, including in the infrared with Spitzer
  \citep[][ -- \#1ab]{2005MNRAS.364..443E,2006astro.ph..4554Y}.  GLARE
  1054 was identified from the $i'$-drop selection in the HUDF, GLARE
  3001 and 3011 from the somewhat brighter GOODS selection. GLARE 1008
  is a source selected from the initial early data release list of red
  sources in the HUDF survey. It lies outside the final overlap region
  of the HUDF which has been used for catalogue construction and
  analysis, but within the noisy outer regions of the HUDF mosaic. It
  is technically below the detection limit of the GOODS survey,
  although it is faintly detected in the GOODS $z'$-band. It is
  faintly detected in the $z'$-band in the shallower edges of the HUDF.

 Given the higher resolution and more sensitive
  flux limit of this spectroscopy, we can rule out a low redshift
  interpretation for these sources and confirm them as $z\approx6$
  Lyman-$\alpha$ emitters. The spectral resolution of the GMOS
  configuration used in our 2004 GLARE mask is
  sufficient to resolve the [OII]
  $\lambda_\mathrm{rest}=3727,3729$\AA\ doublet (and does so in the
  case of the two known [OII] emitters on our 2004 GLARE mask), and we would expect
  to identify weaker emission lines such as [NII], [SII] within our observed
  redshift range if the detected line was H$\alpha$
  ($\lambda_\mathrm{rest}$=6563\AA). The other strong optical
  emission lines (H$\beta$\,4861, [OIII]\,4959,5007) would always
  yield other strong lines within our spectral range. Sources at these
  low redshifts are also unlikely to satisfy the colour cut criterion
  used for target selection.
  
  All five detected strong Lyman-$\alpha$ emission lines also show
  significant asymmetry as expected for high redshift emitters (in
  which the blue wing of the line is significantly self absorbed, and
  the red wing broadened by re-emission). This phenomenon is well
  known at $z\approx3$ and is believed to arise in outflows from the
  actively star-forming galaxies \citep{2003ApJ...584...45A}. Similar
  asymmetry has now been observed in all $z>5.5$ galaxies confirmed to
  date by spectroscopy that resolves the Lyman-$\alpha$ emission line
  \citep[e.g.][]{2003MNRAS.342L..47B}, suggesting that similar
  outflows are produced at higher redshift galaxies.

\begin{table*}
\begin{center}
\begin{tabular}{llccccccc}
ID   & RA \& Declination & Alternate & z & $z'_{AB}$    & $i'-z'$ & Line Flux                & W$_{Ly\alpha}^\mathrm{rest}$ & $L_{Ly\alpha}$\\
     &  J(2000)          & ID        &   &              &         & erg\,cm$^{-2}$\,s$^{-1}$ &   \AA   & $10^{42}$\,erg\,s$^{-1}$     \\
\hline\hline  
1042$^\dag$ & 03 32 40.01 -27 48 14.9 & 20104 & 5.83 & 25.35$\pm$0.02 & 1.64$\pm$0.04 &  1.58 $\times10^{-17}$ &  22.9      & 5.90 \\
1054 & 03 32 33.20 -27 46 43.3 & 42414 & 5.93 & 27.65$\pm$0.07 & 1.45$\pm$0.17 &  0.68 $\times10^{-17}$ &  120     & 2.63   \\
1008 & 03 32 47.97 -27 47 05.1 &       & 6.13 & 28.51$\pm$0.18 & $>$1.35       &  0.43 $\times10^{-17}$ &  159  & 1.80    \\
3001 & 03 32 46.04 -27 49 29.7 &       & 5.79 & 26.69$\pm$0.06 & 1.67$\pm$0.20 &  0.77 $\times10^{-17}$ &  44.1    & 2.83   \\
3011 & 03 32 43.18 -27 45 17.6 &       & 5.93 & 27.47$\pm$0.12 & 1.86$\pm$0.50 &  1.13 $\times10^{-17}$ &  242     & 4.39   \\
\end{tabular}
\end{center}

 $^\dag$ GLARE 1042 has a panoply of alternate names. It is 
  SBM03\#3 in \citet{2003MNRAS.342..439S}, 20104 in 
  \citep{2004MNRAS.355..374B} and is SiD0002, the $z=5.83$ line emitter 
  of \citet{2004ApJ...600L..99D}.

\caption{Strong line emitters on the 2004 GLARE mask. Errors on 
  line fluxes and equivalent widths are approximately 20\%.
  Equivalent widths are calculated from broadband magnitudes,
 accounting for line contamination and Lyman-$\alpha$ forest blanketing.
  2\,$\sigma$ limits on magnitudes are given where appropriate.
  Alternate ID is taken from \citet{2004MNRAS.355..374B}. Note that 
  GLARE 1042 lies in the chip gap of one of the three wavelength
  settings observed, and hence in a region with slightly lower
   signal to noise.}
\label{tab:strong}
\end{table*}

%\hspace{12pt}
%
%OII emitters:\\
%\begin{tabular}{ll}
%Slit   &     Redshift\\
%\hline\hline  
%5008   &       1.22338\\
%5014   &       1.03465\\
%\end{tabular}

  \subsection{Possible Emission Lines}
  \label{sec:maybe}
 
\begin{figure*}
\begin{center}
\begin{tabular}{ccc}
\resizebox{0.55\columnwidth}{!}{\includegraphics{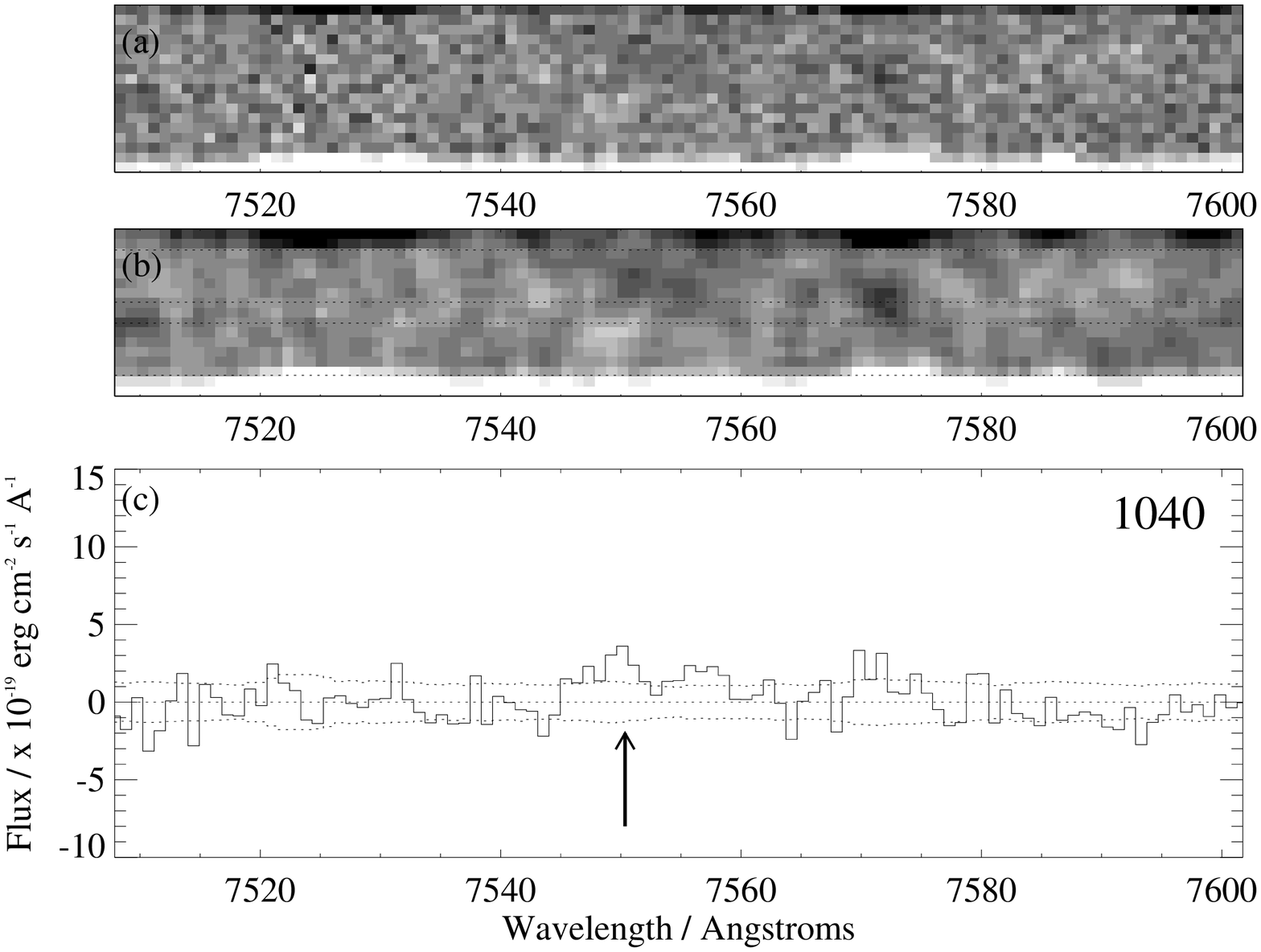}} &
\resizebox{0.55\columnwidth}{!}{\includegraphics{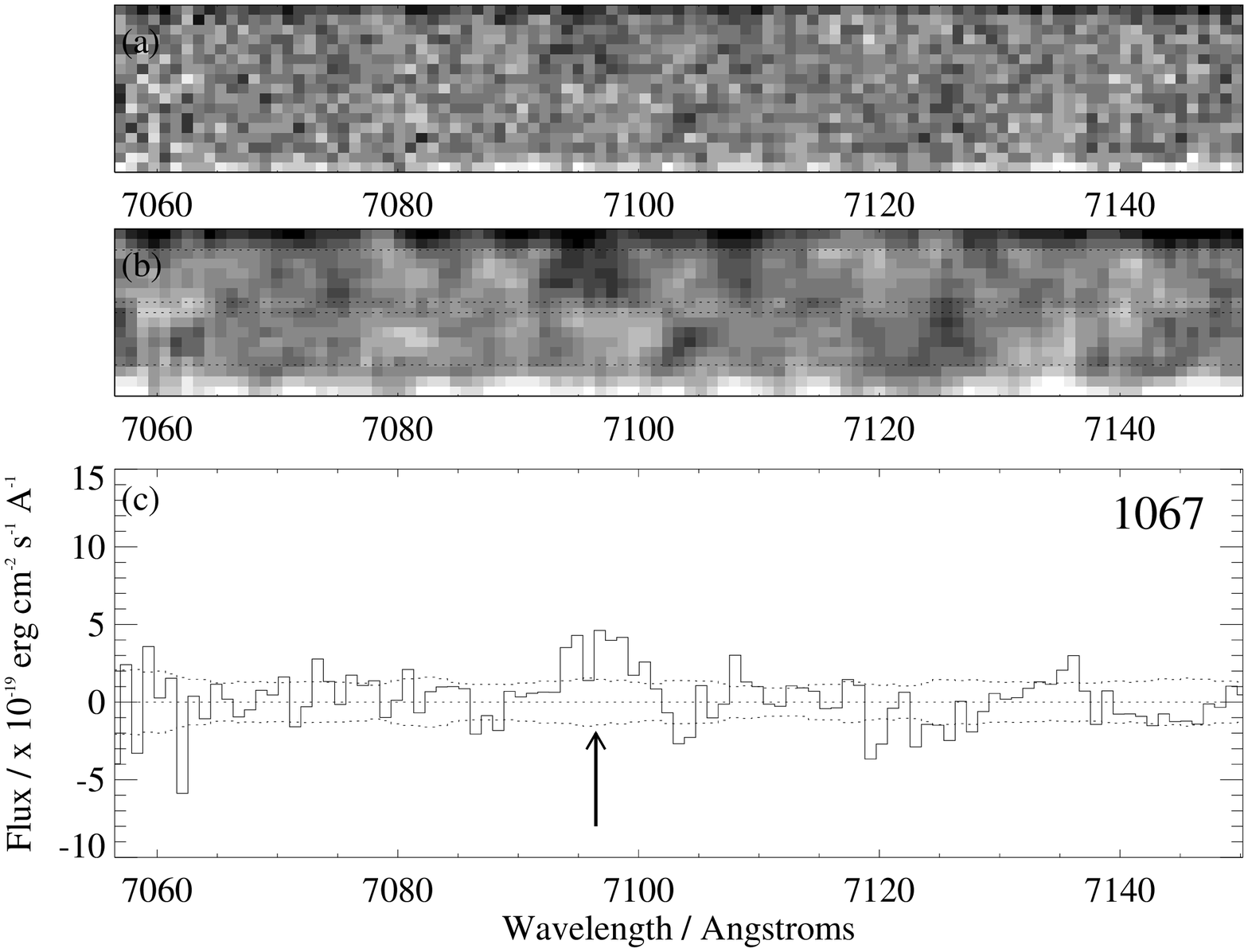}} &
\resizebox{0.55\columnwidth}{!}{\includegraphics{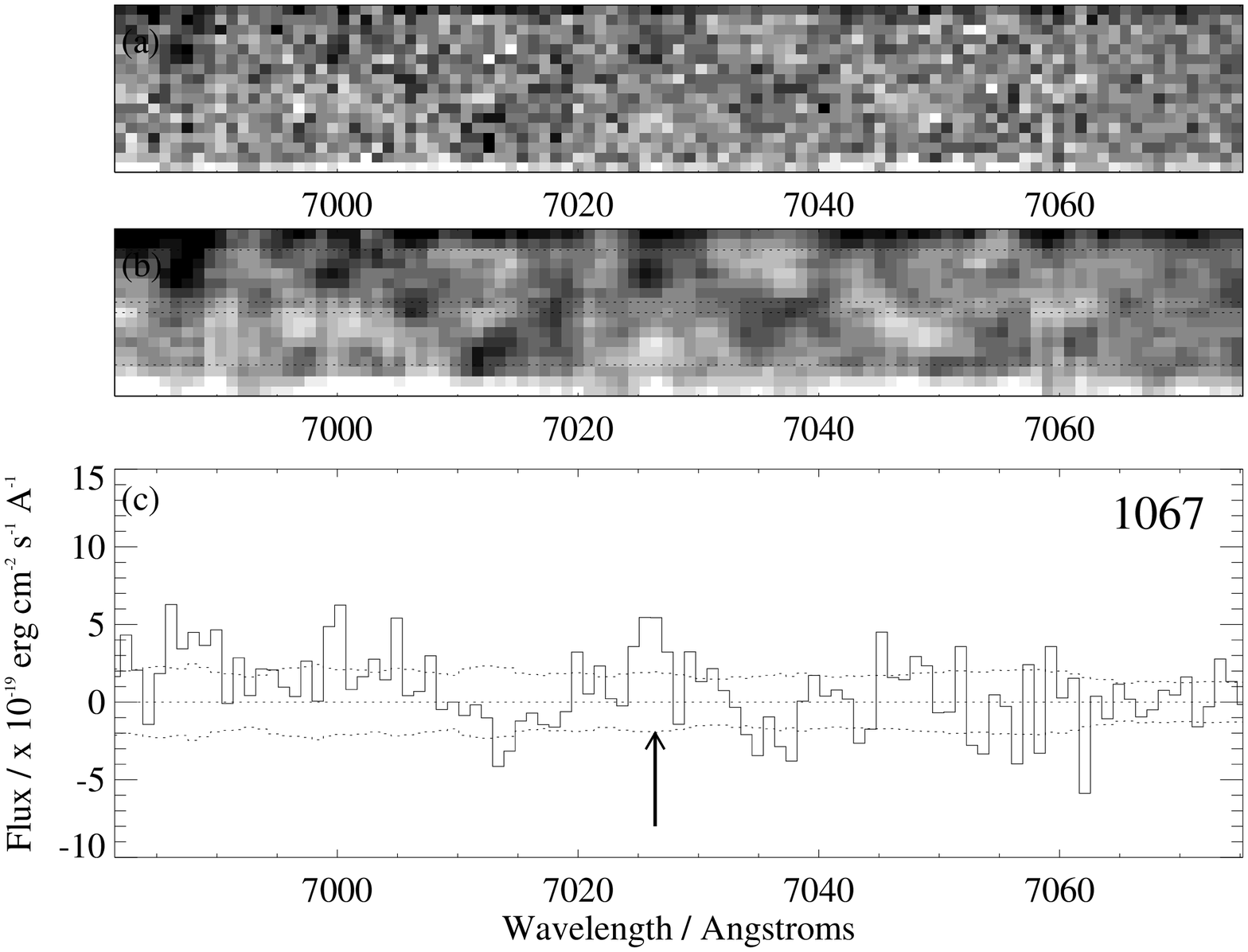}} \\
\resizebox{0.55\columnwidth}{!}{\includegraphics{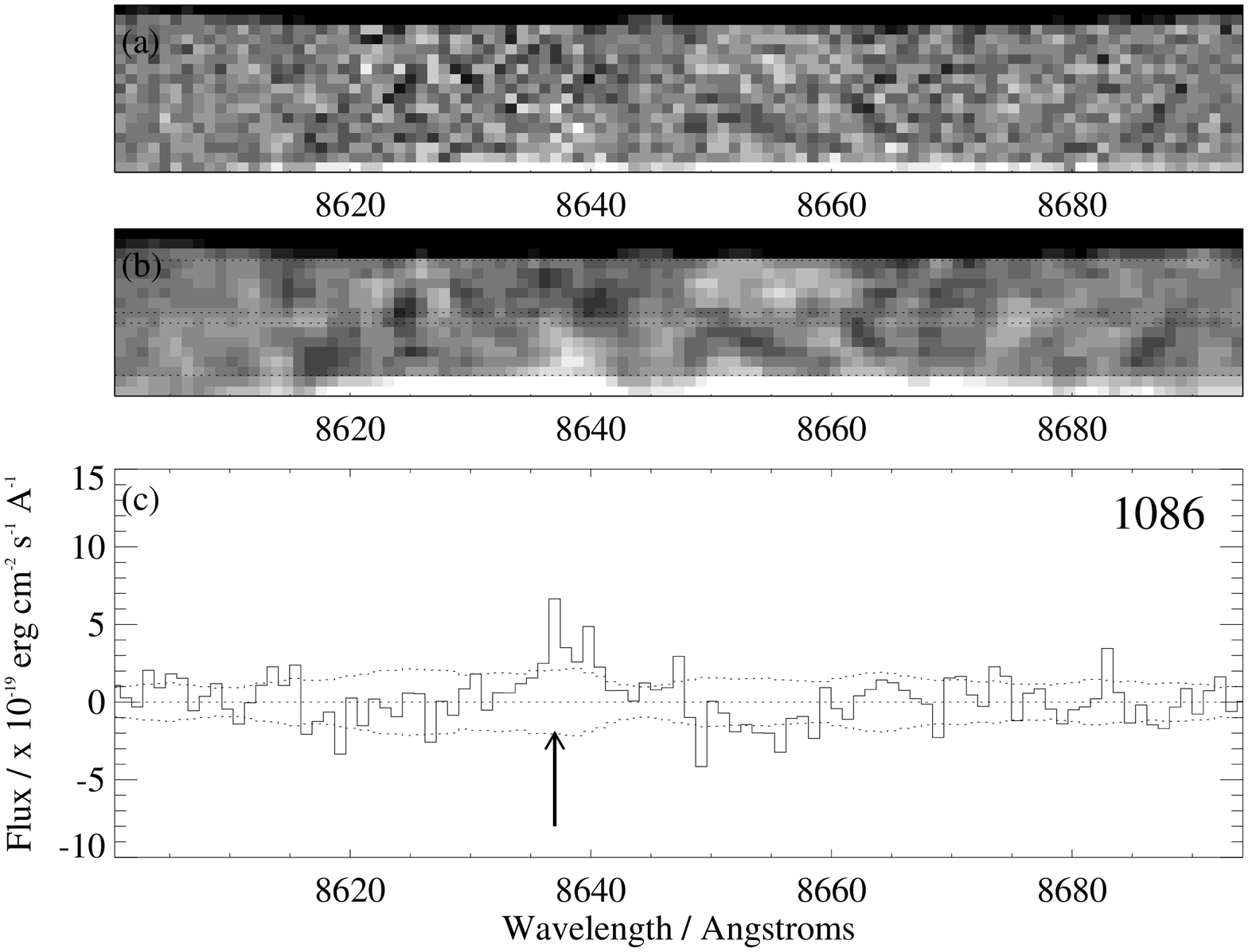}} &
\resizebox{0.55\columnwidth}{!}{\includegraphics{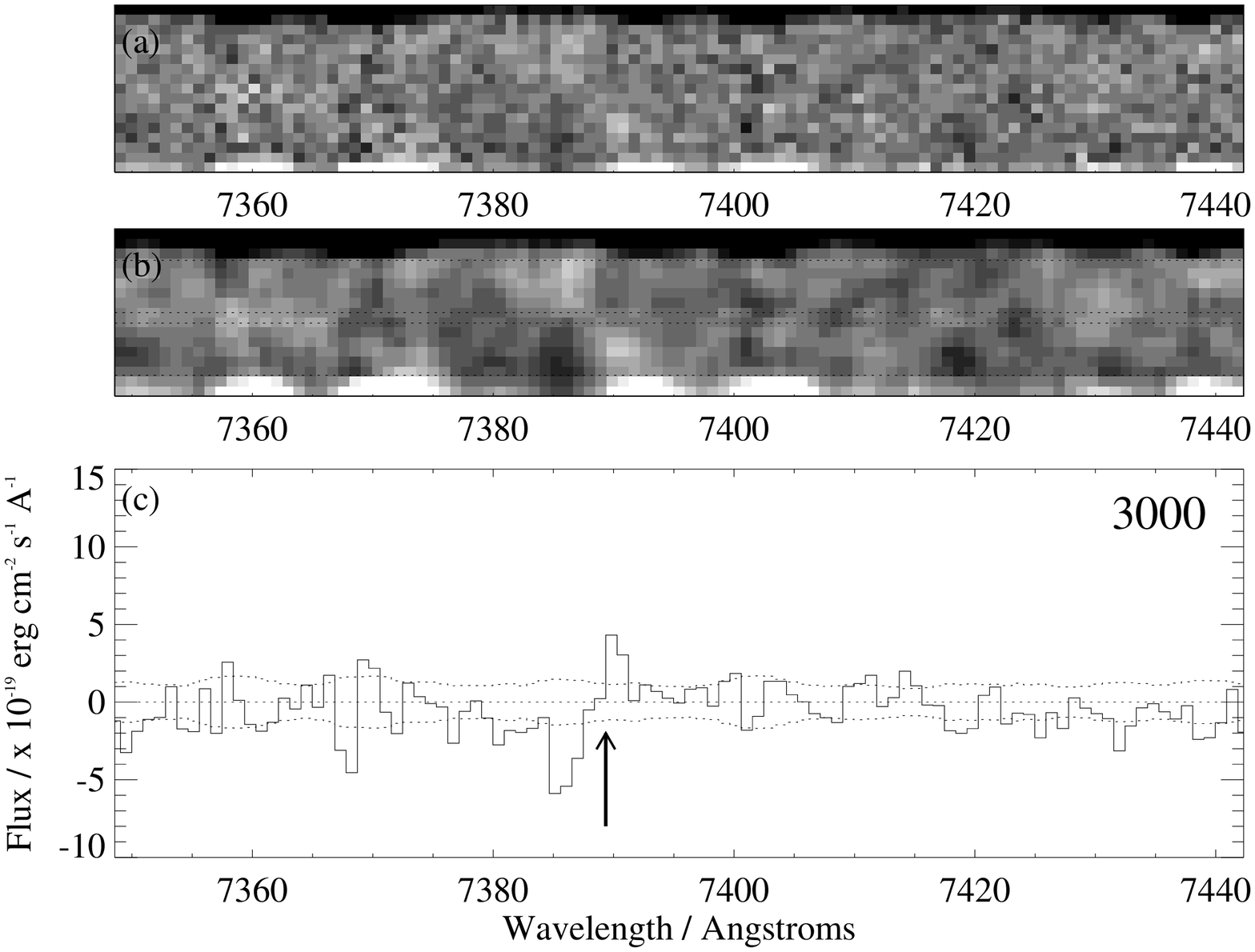}} & \\
\end{tabular}
\end{center}
\caption{One dimensional combined spectrum extracted from both channels
  of the image, and two dimensional spectra, of the possible emission
  line candidates. As in figure \ref{fig:strong}.}\label{fig:possible}
\end{figure*}

  We identify a further four sources (listed in table
  \ref{tab:possible}) as possible line emitters. In most cases, the
  detection in each channel is of low significance, but the
  coincidence of positive and negative signals suggests that the
  emission lines are real. Alternately some candidate lines may lie on
  top of skylines, leading to the suspicion that these represent
  skyline residuals. These sources are illustrated in figure
  \ref{fig:possible}. All but one of the candidate emission lines lie
  shortward of the lower redshift limit selected by the $i'$-drop
  technique. Sources at these redshifts would be expected in a
  $v$-drop rather than $i'$-drop selection. It is possible for sources
  with large errors on their $i'-z'$ colour to scatter into the
  $i'$-drop selection, although it is unlikely that photometric
  scatter alone could explain this discrepancy.
  
  Three of these sources were identified in the HUDF $i'$-drop sample,
  one (GLARE 3000) from the GOODS sample. GLARE 3000 was first
  identified as an $i'$-drop source in \citet[][,
  SBM03\#05]{2003MNRAS.342..439S} but considered likely to be an M or
  L class Galactic dwarf star, as it is unresolved in {\em HST} imaging. 
  The candidate emission line in this source
  lies beside a strong sky line residual. FORS2 spectroscopy of this
  source by \citet{2004astro.ph..6591V} also interpreted the spectrum obtained
  as that of a Galactic star. The stellar identification is almost
  certainly correct, given the broadband colours and unresolved half
  light radius of this source, suggesting that possible emission lines
  at this significance should be considered highly suspect.
  
  There are two possible emission lines of similar strength in GLARE
  1067 (at 7037\AA\ \& 7099\AA), rendering it unlikely that this
  source lies at high redshift.  The observed line separation may be
  consistent with [OIII] emission
  ($\lambda_\mathrm{rest}=4959,5007$\AA) at $z=0.418$, although a
  galaxy at this redshift is predicted to have colours that are
  significantly bluer (i.e. $i'-z'<0.5$ and detected in the $b$ band,
  as opposed to the observed $i'-z'=1.4\pm0.2$ and no $b$ detection).
  In addition, the redward line of this [OIII] doublet is expected to
  be three times stronger than the blueward line, while the observed
  emission peaks are of comparable strength.  An alternate explanation
  might be that two galaxies, separated by $3000$\,km\,s$^{-1}$ lie
  within the slit. While this source has a close neighbour (at
  $\alpha_\mathrm{J2000}=$03:32:35.8,
  $\delta_\mathrm{J2000}=$-27:48:49, with a separation of $<1$
  arcsecond), the neighbouring galaxy is blue in colour ($i'-z'=0.2$),
  and likely lies at significantly lower redshift.  Therefore the most
  probable interpretation of these lines (if they are real) is that
  they represent [OIII] emission arising not in the targeted galaxy
  but rather in the neighbouring source.
 
  GLARE 1040 and GLARE 1086 are the most convincing candidates in this
  category. GLARE 1040, is an isolated source that clearly drops
  between the $i'$ and $z'$-bands, and is undetected in the HUDF
  $v$-band.  The candidate emission line lies at $z=5.2$, just
  shortward of the $i'$-drop selection, although it is possible for a
  faint galaxy such as this to scatter upwards into the selection
  window.
  
  The candidate line in the final source, GLARE 1086, lies firmly in
  the redshift range selected by the $i'$-drop technique, and may well
  be a high redshift emission line, although the detection is too weak
  to rule out a low redshift [OII] explanation, or to detect line
  asymmetry.

\begin{table*}
\begin{tabular}{llccccccc}
Slit & RA \& Declination & Alternate &  $z$  &  $z'_{AB}$     & $i'-z'$ & Line Flux               & W$_{Ly\alpha}^\mathrm{rest}$ & $L_{Ly\alpha}$\\
     &  J(2000)          &   ID      &     &                &         &erg\,cm$^{-2}$\,s$^{-1}$&  \AA  & $10^{42}$\,erg\,s$^{-1}$\\
\hline\hline  
1040 &  03 32 38.28  -27 47 51.3 & 24458   & 5.21?             & 27.51$\pm$0.07 & 1.60$\pm$0.17 & 0.17$\times10^{-17}$  &  20.2 & 0.49\\
1067 &  03 32 35.83  -27 48 48.9 & 14210   & 4.84?$^{\dagger}$ & 28.08$\pm$0.10 & 1.43$\pm$0.24 & 0.30$\times10^{-17}$  &  81.5 & 0.72\\
     &                           &         & 4.78?$^{\dagger}$ &                &               & 0.31$\times10^{-17}$  &  88.0 & 0.73\\
1086 &  03 32 30.14  -27 47 28.4 & 30359   & 6.10?           & 28.13$\pm$0.11   & 1.46$\pm$0.25 & 0.37$\times10^{-17}$  &  68.1 & 1.53\\
3000$^{\ast}$ &  03 32 38.80  -27 49 53.7 &         & 5.08?  & 25.65$\pm$0.03   &  $>$3.50      & 0.11$\times10^{-17}$  &  2.24 & 0.30\\
\end{tabular}

$^{\dagger}$ GLARE 1067 has two emission lines, which are consistent with [OIII] at $z=0.418$;
if these are instead Lyman-$\alpha$, both are at $z<5$.\\
$^{\ast}$ GLARE 3000 is most likely a low-mass Galactic star.

\caption{Possible line emitters on the 2004 GLARE mask. As in table \ref{tab:strong}.
 The redshift assumes that the emission detected is the Lyman-$\alpha$ line at
 $\lambda_\mathrm{rest}=1215.7$\AA.}
\label{tab:possible}
\end{table*}

\begin{figure*}
\begin{center}
\begin{tabular}{ccc}
\resizebox{0.55\columnwidth}{!}{\includegraphics{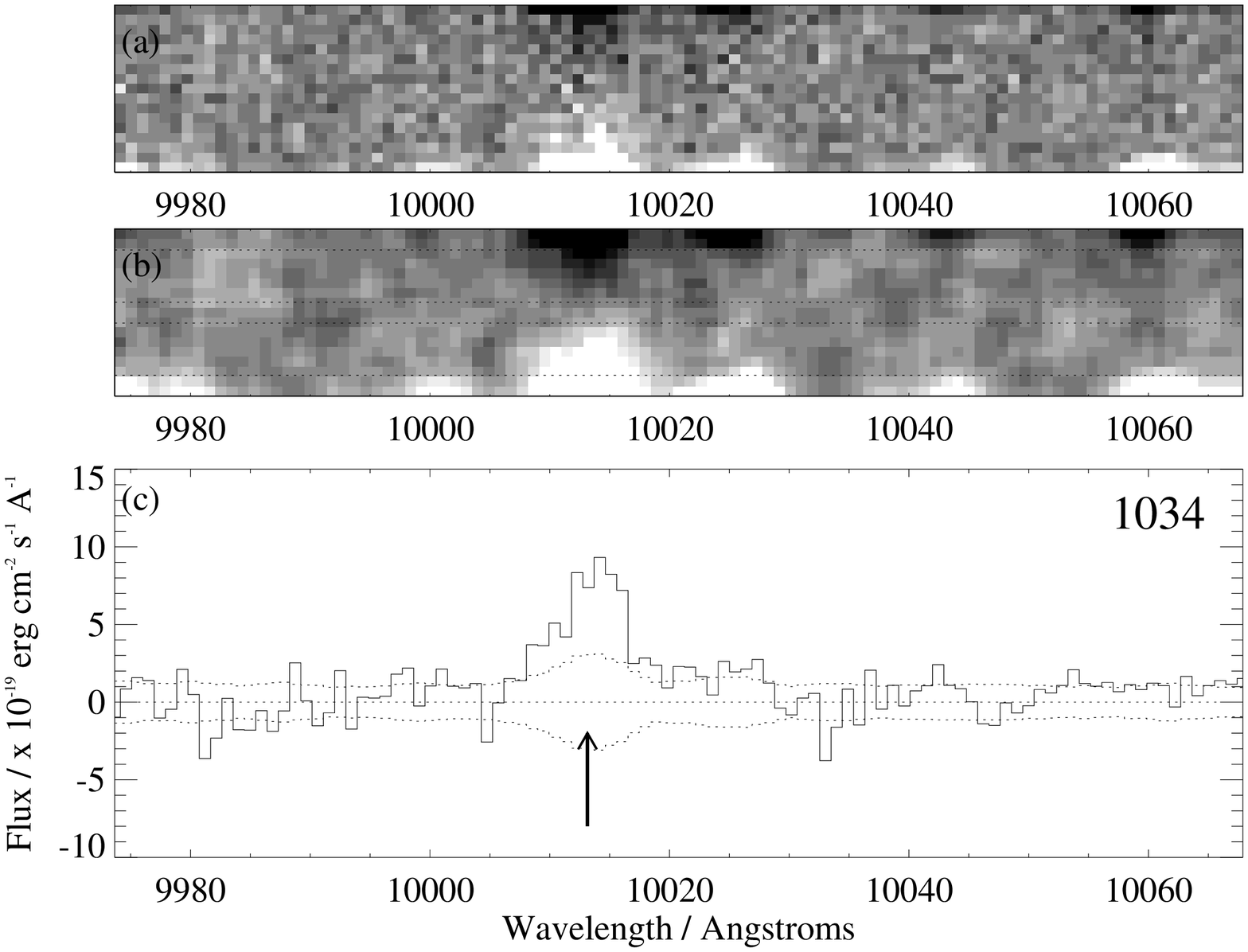}} &
\resizebox{0.55\columnwidth}{!}{\includegraphics{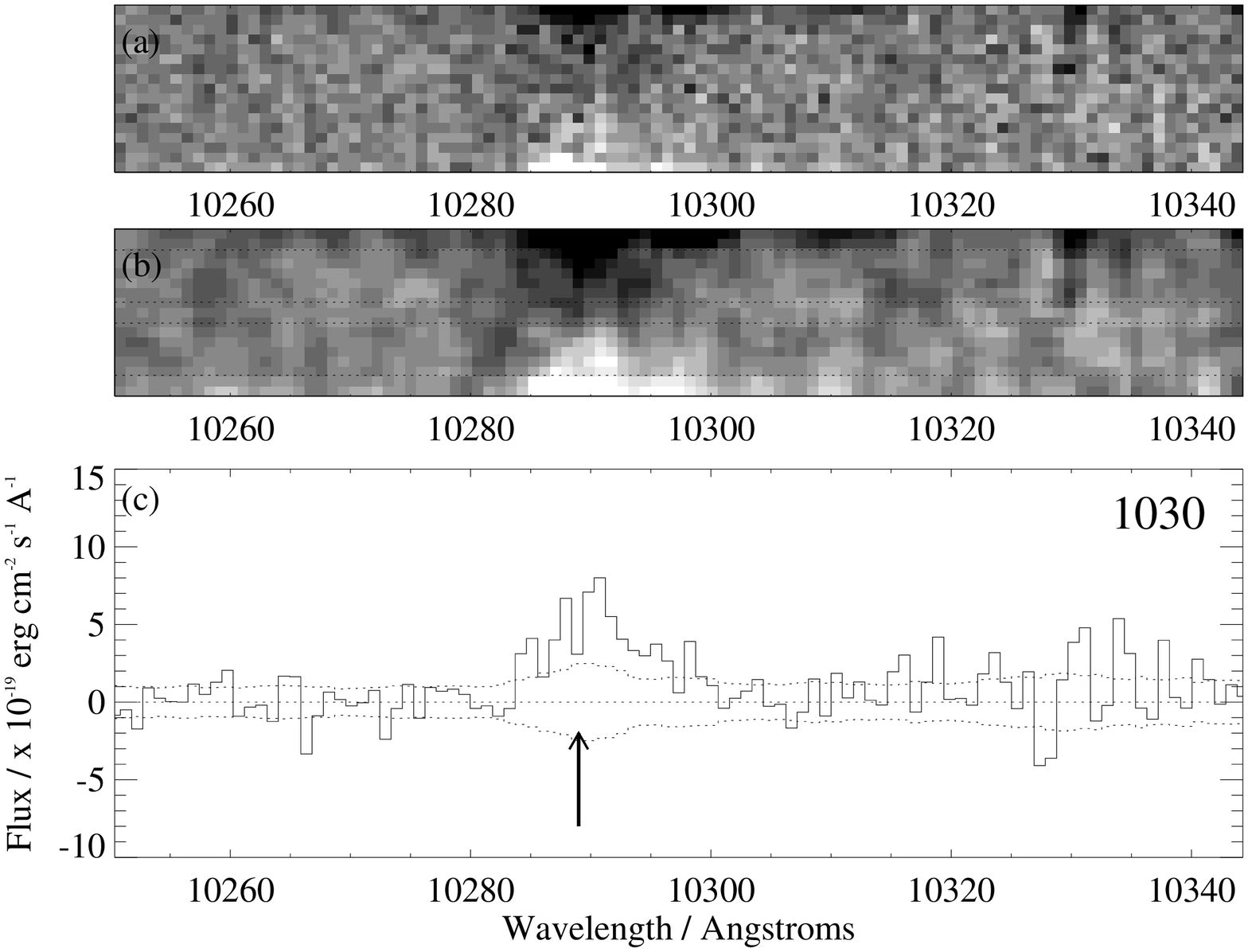}} &
\resizebox{0.55\columnwidth}{!}{\includegraphics{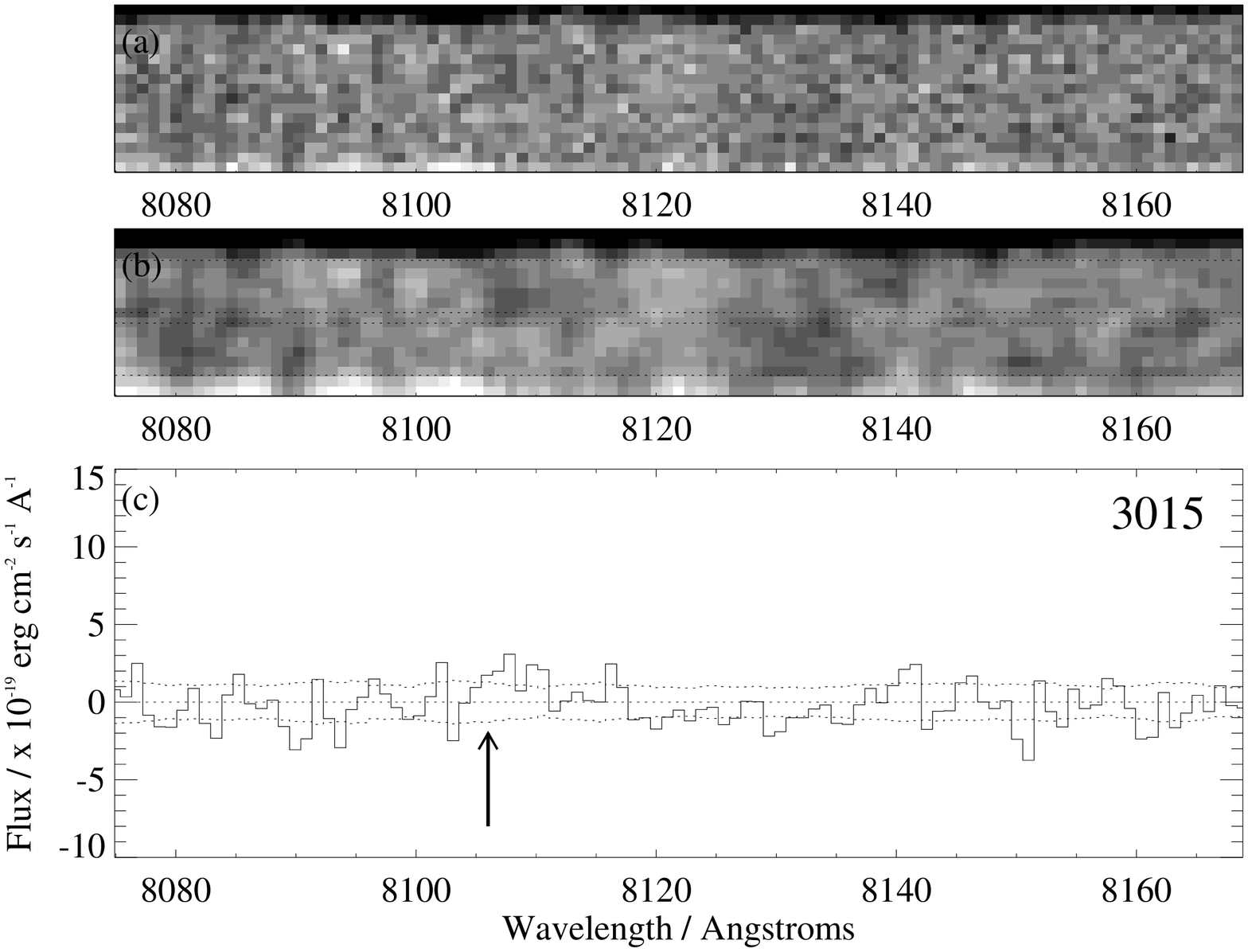}} \\
\resizebox{0.55\columnwidth}{!}{\includegraphics{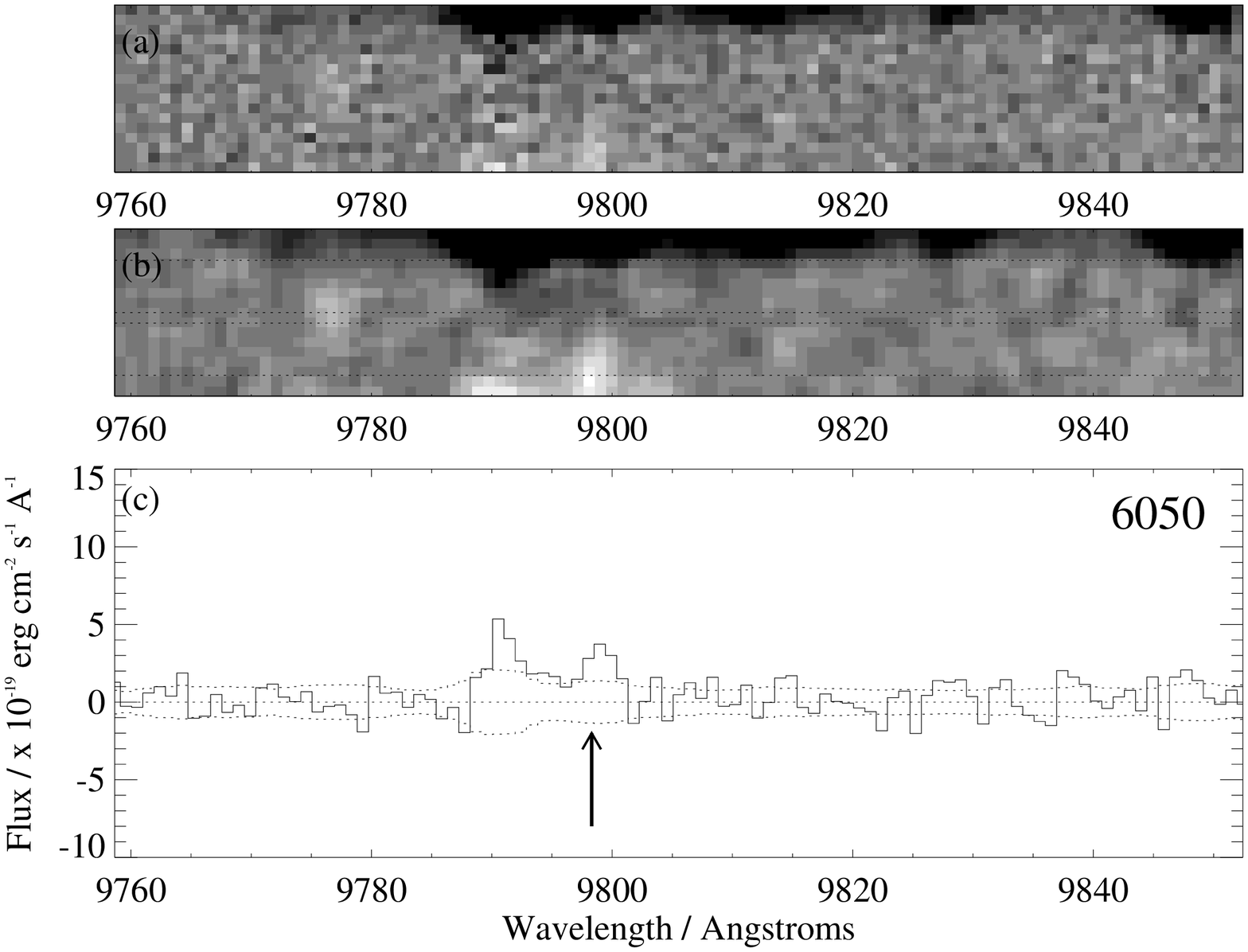}} &
\resizebox{0.55\columnwidth}{!}{\includegraphics{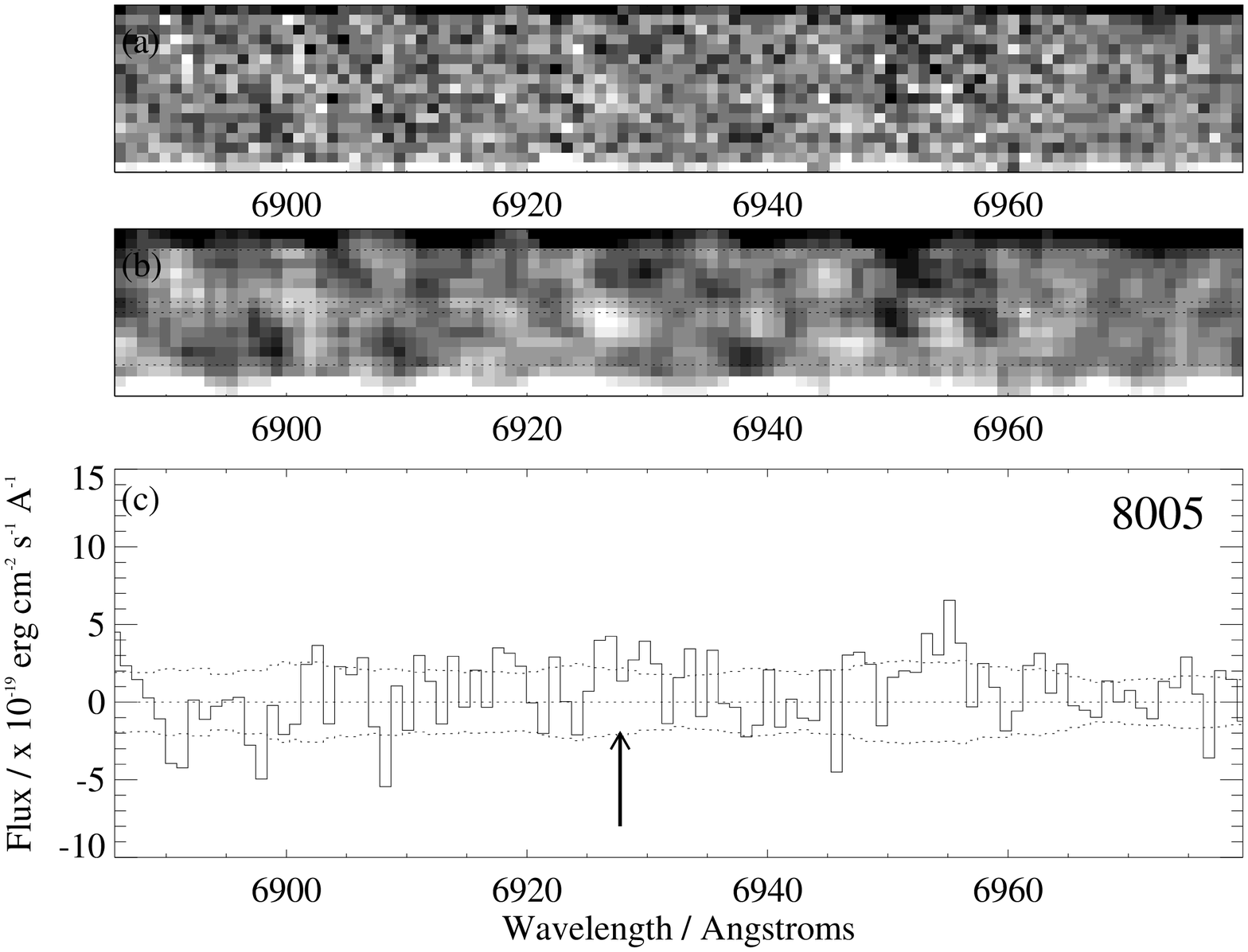}} & \\
\end{tabular}
\end{center}
\caption{One and two dimensional spectra of the `unlikely' 
  emission line candidates presented in table \ref{tab:not}. As in
  figure \ref{fig:strong}.}\label{fig:not}
\end{figure*}

Finally, we identify signals in a further five sources as `unlikely'
emission lines. These are shown in table \ref{tab:not} and figure
\ref{fig:not}. Although the dipole signal from the first two sources
appears strong, they lie on bright skylines and so may be partly or
wholly due to sky subtraction residuals. They may also arrise as a
result of charge diffusion from the adjacent slit. Both candidates lie
at $\lambda>10000$\AA\ (not far from the 10500\AA\ Si bandgap) where
the diffusion is strong. The signal in the remaining candidates is
weak. Slit GLARE 6050 was a blank sky slit, and there is nothing
visible in any waveband at the depth of the GOODS imaging of this
region. The possible serendipitous line emitter is offset from the
centre of the slit by approximately 0\farcs2 to the north.

\begin{table*}
%Almost certainly not emission lines:\\
\begin{tabular}{llccccc}
Slit   &  RA \& Declination           & Alternate &  z  &  $z'_{AB}$     & $i'-z'$       & W$_{Ly\alpha}^\mathrm{rest}$\\
       &  J(2000)                     &   ID      &     &                &               &  \AA  \\
\hline\hline  
1034   & 03 32 39.454  -27 45 43.42   & 52086   &  7.24? & 27.97$\pm$0.09 & $>$2.43       & 10.6\\
1030   & 03 32 41.184  -27 49 14.81   & 10188   &  7.46? & 27.10$\pm$0.05 & 2.04$\pm$0.16 & 4.4\\
3015   & 03 32 27.910  -27 49 41.98   &         &  5.67? & 27.30$\pm$0.10 & $>$1.85       & 16.7\\
6050$^{\ast}$   & 03 32 37.450  -27 49 47.60   &         &  7.05? &         -      &     -         & $>$104 \\
8005   & 03 32 34.392  -27 45 33.01   &         &  4.69? & 27.11$\pm$0.16 & 1.15$\pm$0.36 & 66.6 \\
% 8005 24_2102  
\end{tabular}

$^{\ast}$ 6050 is a `blank sky' slitlet
\caption{The five emission line candidates classified as `unlikely' to be real.
  A limit has been placed on the possible serendipitous emitter GLARE 6050 based
  on non-detection in the $z'$ band to our imaging depth of $z'_{AB}=28.5$. Column
  headings as in table \ref{tab:strong}}
\label{tab:not}
\end{table*}

  \subsection{Sources with No Line Emission}
  \label{sec:limits}
  
  The remaining 21 science targets on the slitmask showed no evidence for line
  emission in the wavelength range observed,
  $\lambda_\mathrm{obs}\approx7800-10000$\AA\ (corresponding to
  $5.4<z<7.2$, with some slit to slit variation depending on slit
  location). 
  
  Of these, 11 formed part of our high priority $i'$-drop selection,
  with marginal targets (5 sources) and sky slits (5 slits)
  constituting the remainder. 
  %While the non-detection of
  %Lyman-$\alpha$ flux decreases the likelihood that these sources lie
  %at high redshift, it does not dismiss it.
  We discuss the implications of these non-detections of line emission below.
  
  It is possible that, despite optimising our experimental setup
  for sky line subtraction, we are missing flux from emission lines
  that fall in regions dense with skylines. The noise in such areas is
  greater than the slit average, making identification of line
  candidates more difficult. Some 35\% 
 % Note to self - checked on sky spectra (where skylines are >1.5 cont)
  of the wavelength coverage of this spectroscopy lies under sky lines
  (defined as regions where sky emission exceeds 150\% of the
  inter-line sky continuum level) which leave Poisson noise and sky
  brightness fluctuation residuals of varying strength, although only
  10\% lies under strong skylines (sky line emission $>5$ times sky
  continuum). The noise under the residuals of bright sky lines is a
  factor of 2.5 greater than that between lines, and might also
  be prone to systematics in the sky subtraction leading to potentially
  spurious emission lines.
 
  Only half of Lyman Break Galaxies at $z\approx3$ show Lyman-$\alpha$
  in emission \citep{2003ApJ...588...65S}, and one quarter have
  W$_{Ly\alpha}^\mathrm{rest}>$20\AA.  Given that we reach a limiting
  rest-frame equivalent width of $\le$6\AA\ for the majority of our
  targets, we might na\"\i vely expect to observe emission in half our
  slits, less ten percent lost to strong sky line residuals.  However,
  the fraction of Lyman Break Galaxies with emission at $z\approx6$ is
  still unknown, and it is likely that a large fraction of $z\approx6$
  $i'$-drop galaxies will not be spectroscopically confirmed until the
  continuum level is reached with either longer exposures or more
  sensitive telescopes, and interstellar absorption lines can be used
  for redshift determination \citep[see, for example,][for rest-frame
  UV continuum spectroscopy of a bright $z=5.5$
  galaxy]{2005ApJ...630L.137D}.
  
  Finally in the case of the marginal candidates used to fill the
  slitmask, it is possible that the true redshift of the sources lies
  above $z=4.5$ (set by the dual requirements of non-detection in the
  $b$ band, and the $i'-z'$ colours of these sources), but below the
  limit of our spectroscopy.  Two of the slits without detections were
  placed on $v$-drop galaxies expected to lie in this range, while a
  further four slits were placed on sources slightly too blue to meet
  our strict selection criteria.  It seems unlikely that
  photometric scatter could place these sources in our detectable
  redshift range.

We will discuss the equivalent width limits on our $i'$-drop
galaxies in Section~\ref{sec:EW}.
%  For a discussion of the equivalent width limits of these sources,
%  and in particular those satisfying our $i'$-drop selection, 
%  please see section \ref{sec:EW}.

%On the Mask and not in the 'strong' or 'possible categories:
%
%1001,1004,1009,1022,1030,1034,1045,1047,1060,1077 -  (UDF z=6 candidates)
%
%3002,3033 - (High priority GOODS z=6 cands)
%
%3015,3030 -  (Normal GOODS z=6 cands)
%
%8001,8005,8015 - (Slightly Fainter or Bluer GOODS z=6 cands)
%
%1061 - (From the UDF list but not z'<28.5, i-z>1.3)
%
%%4069,4087 -  (Bremer GOODS z=5 cands)
%
%6050,6051,6052,6060,6061,6062 -  (Blank sky slits)

%1022  -     03 32 32.722  -27 46 37.24
%1061  -     03 32 30.516  -27 47 20.11
%4069   03 32 41.345  -27 48 43.09
%4087   03 32 48.499  -27 49 34.79
%6051   03 32 39.401  -27 45 53.21
%6052   03 32 43.130  -27 45 25.31
%6060   03 32 31.510  -27 50 05.89
%6061   03 32 30.351  -27 45 00.50
%6062   03 32 30.900  -27 45 38.49

%8001 24_3499  03 32 37.752  -27 45 05.51 27.40 0.11 1.59 0.37
%8015 23_1296  03 32 46.536  -27 49 58.40 27.33 0.17   >1.78 

  \subsection{Agreement with Other Spectroscopic Surveys}
   \label{sec:others}
   
   A subset of sources in our sample have also been observed
   spectroscopically as part of the concerted campaign of follow up
   observations to the GOODS survey.

   Sources in the GOODS-S field have been targeted for 8m spectroscopy
   by FORS\,2 on the VLT (Vanzella et al, 2006), by DEIMOS on Keck
   \citep[][, Bunker et al 2006]{2004ApJ...607..704S} and by GMOS on
   Gemini (Paper I and this work). In addition, this field was
   surveyed with the {\em HST}/ACS grism as part of the GRAPES survey
   \citep{2005apj_malhotra}. This slitless spectroscopy has been
   obtained to varying depths, and at different spectral resolutions.
   In particular, the GRAPES grism spectroscopy is of too low a
   resolution to detect all but the highest equivalent width
   Lyman-$\alpha$ emission lines adjacent to the Lyman-$\alpha$ forest
   continuum break (compare, for example, the DEIMOS spectrum of
   SBM03\#1/GLARE1042 in Stanway et al.\ 2004 showing the strong
   Lyman-$\alpha$ line, with the {\em HST}/ACS spectrum of the same
   object, SiD002, in fig.~2a of Dickinson et al.\ 2004, which shows a
   continuum break but the line is washed out).  However, GRAPES does
   provide a redshift estimate by localising the wavelength of the
   Lyman-$\alpha$ break in the spectrum.
   
   Where redshifts have been determined by multiple groups, our
   results are generally in reasonable agreement with previous
   observations, given the difficulty in obtaining precise redshifts
   from the self-absorbed and resonantly-broadened Lyman-$\alpha$
   lines at moderate spectral resolution. In the case of GLARE 1042,
   our measured redshift of $z=5.83$ is in close agreement with that
   measured by FORS\,2 \citep[$z=5.82$,][]{2004astro.ph..6591V}, by
   the GRAPES team \citep[$z=5.8$,][]{2005apj_malhotra} and, using
   Keck/DEIMOS spectra, by \citet[][ $z=5.83$]{2004ApJ...607..704S}.
   GLARE 3001 was spectroscopically identified by FORS\,2 as a line
   emitter at $z=5.78$, agreeing with our redshift of $z=5.79$.
   
   GLARE 1054 and 1030 were both placed at $z=5.7$ by GRAPES
   spectroscopy \citep{2005apj_malhotra}. This contrasts with our
   measured redshifts of $z=5.9$ for GLARE 1054 and (tentatively)
   $z=7.4$ for GLARE 1030. The discrepancy for GLARE 1054 is within
   the expected level of agreement for GRAPES grism spectroscopy and
   so these results are consistent.  Given the low significance of our
   line candidate in GLARE 1030, and the $i'$-drop redshift selection
   function, the GRAPES redshift remains the more likely.
   
   The remaining source, GLARE 3000 was identified by the VLT/FORS\,2
   observations as a Galactic star on the basis of weakly detected
   [OI] and [NII] lines. This source, which corresponds to the
   unresolved candidate SBM03\#5 in \citet{2003MNRAS.342..439S}, falls
   in our `possible' category. We note that the FORS II spectrum is
   flagged with their quality class `C' and that there is no secure
   continuum detection.  Despite this, a Galactic star is still the
   most likely identification of this target, illustrating the caution
   with which we present our fainter line candidates.
%However we note the discrepancy between these two results for
%   further study.
% Note - we have a DEIMOS spectrum of this source (November 2005, 041) with a candidate line on it. at a slightly different redshift. Need to crosscheck all this spectroscopy.

   No other target on our 2004 GLARE mask has been reported as a line
   emitter by other teams, or has an estimated redshift from GRAPES
   spectroscopy. 

\subsection{Limits on NV and other Emission Lines} 
  \label{sec:other_lines}

%1042: Possible line at 8345\AA - 1221\AA (rest), may be skyline residual.
  
  Lyman break galaxies at $z\approx3$ show few other emission lines in
  the rest frame ultraviolet. The composite spectrum of $\approx$1000
  such galaxies produced by \citet{2003ApJ...588...65S} shows weak
  emission features due to SiII* ($\lambda_\mathrm{rest}=$1265\AA,
  1309\AA, 1533\AA), OIII] ($\lambda=$ 1661\AA, 1666\AA) and CIII]
  ($\lambda=$1908\AA), and absorption features due to stellar winds,
  primarily SiII and CIV.

The presence of an AGN in our target galaxies could also lead to the
presence of emission lines due to high excitation states, primarily NV
($\lambda_\mathrm{rest}=$1238.8\AA, 1242.8\AA) and SiIV
($\lambda=$1394\AA, 1403\AA), and rarely OV] ($\lambda=$1218\AA).

For a galaxy at $z=6$, our spectra extend to rest frame wavelengths of
$\approx$1428\AA. Given that we are not able to measure a high signal
to noise continuum on any one target, we are unable to measure
absorption features in the spectra, and therefore confine ourselves to
searching for evidence of other emission features in the spectra of
our Lyman-$\alpha$ emitters.

A careful inspection of our five good Lyman-$\alpha$ emission line
candidates does not provide evidence for any other emission lines at
the redshift of Lyman-$\alpha$. While this is not surprising given the
weakness of secondary lines, the failure to detect NV in this spectrscopy
suggests a large Ly-$\alpha$ / NV ratio. Using our 3$\sigma$ limit on
 undetected lines as an upper constraint on NV flux we determine that
$f(\mathrm{Ly}\alpha)/f(\mathrm{NV}) > [10.5, 4.5, 2.8, 5.1, 7.5] (3\sigma)$
respectively for GLARE targets [1042, 1054, 1008, 3001, 3011]
Typical line ratios for AGN are $f(\mathrm{Ly}\alpha)/f(\mathrm{NV})=4.0$
 (Osterbrock 1989), while those for galaxies are greater than this. The 
limits we determine disfavour an AGN origin to the Lyman-$\alpha$ flux in
 all but the faintest target (in which the constraint is too weak to make a
 firm statement). These limits constrain AGN
activity, if present, to only a weak contribution to the rest frame
ultraviolet flux.

\section{The Equivalent Width Distribution of GLARE line Emitters} 
  \label{sec:EW}

\subsection{The Observed Equivalent Width Distribution} 
  \label{sec:EWobs}
  
  Using the mean variance in the background of the exposed slits, and
  the broadband magnitudes of the targeted galaxies, we are able to
  calculate limits on the rest frame equivalent width $W_0$ for those
  sources which satisfy our colour selection criteria and yet are undetected
  in our spectroscopy.  These are presented in table \ref{tab:ewlims}.
  In each case we assume that the galaxy lies at the mean $i'$-drop
  redshift ($z=6.0$) and that the emission line has not been lost behind a
  the subtraction residual of a bright night sky line.
  
\begin{figure}
\begin{center}
\resizebox{0.75\columnwidth}{!}{\includegraphics{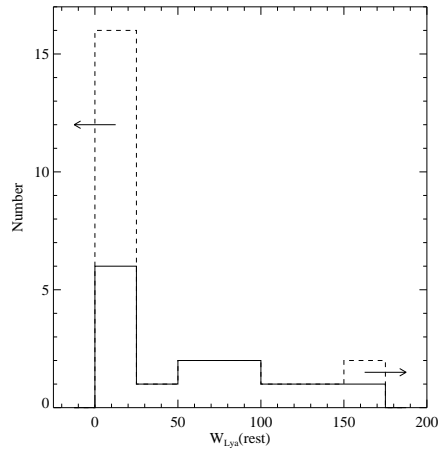}}
\end{center}
\caption{The distribution of rest frame equivalent widths for the
  GLARE mask, divided into bins of 25\AA. Line emission candidates are
  shown with solid lines, sources with equivalent widths $>150$\AA\ are
  shown at 150\AA, and those for which only upper and lower limits are
  available are added to the distribution to produce the dotted
  lines.}\label{fig:ewlims}
\end{figure}

  Combining our line limits with those sources for which emission
  lines have been identified or tentatively proposed forms a sample of
  24 sources uniformly selected from their $i'$-drop colours. The
  resultant distribution of Lyman-$\alpha$ equivalent widths is
  plotted in figure \ref{fig:ewlims}.
  
  If all the sources for which candidate emission lines are identified
  in this paper prove to be $z\approx6$ Lyman-$\alpha$ emitters, then
  the escape fraction of Lyman-$\alpha$ photons at $z\approx6$ appears
  qualitatively similar to that at $z\approx3$.  From our $z\approx 6$
  sample, 66\% of the sources (16 out of 24) have Lyman-$\alpha$
  equivalent widths $<25$\AA, compared with 75\% at $z\approx3$
  \citep[][]{2003ApJ...588...65S}, although the lower redshift sample
  loses a smaller fraction due to skyline contamination. The samples
  probe to similar points on the luminosity function in both cases
  (approximately 0.1\,L$^\ast$). These
  fractions are consistent within the errors on our small number
  statistics.  Harder to explain in comparison with lower redshift
  galaxies is the tail stretching to very high equivalent widths
  ($>200$\AA) observed in this survey, a trait also observed in some
  narrowband selected sources at this redshift
  \citep[e.g.][]{2002ApJ...565L..71M} and in other $i$-dropout Lyman break galaxies \citep[e.g.][ who find one source with $W_0=$150\AA]{2006astro.ph.12454D}.  Although the number statistics
  are small, we observe four line emitters ($17\pm 8$\%) with
  equivalent widths in the range $50-100$\AA, and a further four with
  $W_0>$100\AA\ (three of which come from our `robust' list of line
  emitters).  This contrasts with the Lyman Break Galaxy population at
  $z\approx3$ in which $<5$\% of galaxies have line emission with
  $W_0>100$\AA\ \citep{2003ApJ...588...65S}.
  
  While high redshift galaxies at both redshifts are selected on their
  rest-frame ultraviolet continuum and the spectral break caused by
  Lyman-$\alpha$ absorption, the two populations are not identical.
  
  The sample discussed here reaches some two magnitudes fainter than
  the tentatively proposed and still uncertain typical luminosity
  L$^\ast$ at $z\approx6$
  \citep{2006ApJ...653..53B,2004MNRAS.355..374B}. This contrasts with
  a sample reaching just one magnitude below L$^\ast$ at $z=3$.
  \citet{2003ApJ...588...65S} considered subsamples at $z=3$, dividing
  their spectroscopic data into quartiles based on rest-frame
  equivalent width. They found a modest trend in the strength of line
  emission with magnitude. Galaxies in their highest equivalent width
  quartile are some 0.2 magnitudes fainter on average than their
  quartile of weak to moderate line emission. It is possible that the
  strong line emission observed here is more typical of sources with
  faint rest-frame ultraviolet continuum suggesting that faint sources
  differ physically from brighter members of the population.
  
  A second difference between the samples is intrinsic rather than
  arising from a selection effect. Several authors
  \citep[e.g.][]{2005MNRAS.359.1184S,2006ApJ...653..53B,2003ApJ...593..630L}
  have now observed that Lyman break galaxies at $z>5$ have steeper
  rest-frame ultraviolet slopes than those at $z\approx3$.
  \citet{2006ApJ...653..53B} interpret this as indicating that the
  dust properties of this population evolve over redshift. While a
  steep rest-ultraviolet slope can also arise due to low metallicity
  or a top-heavy initial mass function (as discussed below) dust
  evolution is a natural interpretation. At $z\approx6$ the universe
  is less than 1 gigayear old and galaxies may not have had time to
  develop a high dust content.  \citet{2003ApJ...588...65S} found that
  $z\approx3$ galaxies with high equivalent widths in Lyman-$\alpha$
  also had lower mean dust extinction. Lyman-$\alpha$ photons are
  resonantly scattered by dust and hence the line is preferentially
  absorbed with respect to the rest-frame ultraviolet continuum. If
  the distribution of Lyman-$\alpha$ equivalent widths in the
  $z\approx3$ population is truncated by dust absorption, this could
  produce an apparent `excess' of strong lines at high redshift.
  However, even zero dust absorption cannot explain equivalent widths
  exceeding 200\AA\ unless the sources are also very young and very
  low in metallicity. The full explanation for the equivalent width
  distribution observed in the GLARE data may well be a combination of
  these effects and those discussed below.
  
  Several other possible explanations exist for both the steepening of
  the rest frame UV slope and the difference in equivalent width
  distributions.  An interpretation of contaminant galaxies at lower
  redshifts seems unlikely due to our $i'$-drop selection criteria;
  low redshift sources with strong spectral breaks are likely to have
  more than one emission line in our observed redshift range.
  
  A high equivalent width line can arise if the observed Lyman-$\alpha$
  photons excited by a population of AGN rather than by young, hot
  stars. The luminosity function of AGN is poorly constrained at these
  magnitudes and redshifts, but the space density of such sources is
  expected to be very low (e.g., the $z>6$ SDSS QSOs, Fan et al.\ 2003). 
  At $z\approx6$, the deep 2\,Ms X-ray
  exposure of the UDF and surrounding region
  \citep{2003AJ....126..539A} would detect only the brightest AGN
  ($L>10^{42}$\,ergs\,s$^{-1}$\,cm$^{-2}$). None of the GLARE targets
  are detected in this X-ray observation. AGN would also be expected to 
  show emission lines that are not just strong but also broad, while 
  none of the GLARE line candidates are broad. There is also an
  absence of high-ionization lines such as NV\,1240\,\AA\ which
  are common in AGN.
  
  The tail of line emitters extending to higher equivalent widths may
  also arise from a difference in the fundamental properties of the
  stellar population between $z\approx6$ and that at lower redshifts.
  
  Modeling of emission from metal-free Population III galaxies
  predicts rest frame Lyman-$\alpha$ equivalent widths $>$1000\AA\ for
  young starbursts ($<$2\,Myr), decreasing to
  W(Ly-$\alpha$)$\sim$500\AA\ for older bursts
  \citep{2002A&A...382...28S}. These very high equivalent widths arise
  from the relatively hard spectrum of metal-free reactions in the
  most massive stars.  However it is unlikely that zero metallicity
  (population III) stars are still contributing significantly to the
  emission from massive stars almost a billion years after the Big
  Bang, particularly given the identification of stellar populations
  $>100$\,Myr old in some $z>5$ sources
  \citep[e.g.][]{2005ApJ...618L...5E,2005MNRAS.364..443E}. Further
  evidence for moderate metallicity at $z>5$ has been observed in the
  spectroscopy of bright AGN from the Sloan Digital Sky Survey. Metals
  including iron \citep{2003ApJ...594L..95B} and carbon
  \citep{2005A&A...440L..51M} have been detected from even the highest
  redshift quasar (at $z=6.4$).
  
  By contrast, even 1/20th solar metallicity leads to a sharp
  reduction in the peak (zero age) equivalent width predicted to
  $\sim$300\AA, with a more typical W(Ly-$\alpha$)$\sim$100\AA\ by an
  age of 10-100\,Myr \citep{2002ApJ...565L..71M}. Most of the
  candidate emission lines presented in this study can hence be
  explained with normal, if low metallicity, populations. However at
  least two of our good line emission candidates have rest frame
  equivalent widths exceeding 200\AA. This is possible if the galaxy
  is in the first few Myrs of an ongoing starburst, but may also
  provide evidence for variation in the initial mass function of star
  formation.
  
  High equivalent widths of Lyman-$\alpha$ emission can arise from a
  ``top-heavy'' initial mass function (i.e. star formation weighted
  towards a population of high mass stars).
  \citet{2004ApJ...617L...5M} calculated the Lyman-$\alpha$ equivalent
  width expected for a metal-enriched population with a very extreme
  IMF, weighted towards massive stars (i.e. IMF slope $\alpha=0.5$).
  As is the case for low metallicity populations, the flux is weighted
  towards a harder spectrum, and Lyman-$\alpha$ emission is
  strengthened. They found that such an IMF could explain line
  equivalent widths of up to 240\AA\ at ages of a few Myr, with higher
  IMFs possible for very young bursts. 
  
  The hard rest-frame ultraviolet spectrum associated with such an IMF
  may also be consistent with the steep rest frame ultraviolet
  spectral slopes observed in $z\approx6$ galaxies
  \citep[e.g.][]{2005MNRAS.359.1184S}. While the evidence from the
  GLARE study is limited, with the number of high equivalent width
  sources small, the existence of such sources suggests that the
  environment of star formation at $z\approx6$ is less metal enhanced,
  or weighted towards more massive stars than that at $z\approx3$.

\begin{table*}
\begin{tabular}{llcccc}
ID   & Alt   &   RA \& Declination         &   $z'_{AB}$ &  $i'-z'$ & W$_{Ly\alpha}^\mathrm{rest}$\,/\,\AA     \\
\hline \hline                                                           
1001 & 48989 &  03 32 41.43  -27 46 01.2 &  28.26$\pm$0.12 & $>$2.1 (2\,$\sigma$)   &  $<$6.5 \\
1004 & 46223 &  03 32 39.86  -27 46 19.1 &  28.03$\pm$0.10 & $>$2.37 (2\,$\sigma$)   &  $<$5.2 \\
1009 & 12988 &  03 32 38.50  -27 48 57.8 &  28.11$\pm$0.11 & $>$2.29 (2\,$\sigma$)    &  $<$5.6 \\
1045 & 21111 &  03 32 42.60  -27 48 08.8 &  28.02$\pm$0.11 & 1.67$\pm$0.26  &  $<$5.2 \\
1047 & 35271 &  03 32 38.79  -27 47 10.8 &  28.44$\pm$0.14 & 1.33$\pm$0.30  &  $<$7.6 \\
1060 & 11370 &  03 32 40.06  -27 49 07.5 &  28.13$\pm$0.11 & $>$2.28 (2\,$\sigma$)    &  $<$5.7 \\
1077 & 16258 &  03 32 36.44  -27 48 34.2 &  27.64$\pm$0.07 & 1.42$\pm$0.16  &  $<$3.7 \\
3002 &       &  03 32 43.35  -27 49 20.4 &  26.89$\pm$0.07 & 1.42$\pm$0.20  &  $<$1.8 \\
3030 &       &  03 32 48.94  -27 46 51.4 &  27.04$\pm$0.08 & 1.41$\pm$0.23  &  $<$2.1 \\ 
3033 &       &  03 32 49.08  -27 46 27.7 &  27.18$\pm$0.09 & 1.40$\pm$0.27  &  $<$2.4 \\
\end{tabular}
\caption{The 2004 GLARE targets for which no line emission was observed.
  Limits on the equivalent width are calculated using the $z'$-band magnitude
 to determine the continuum level, the $3\,\sigma$ standard deviation in
 the background as the minimum line flux and accounting for IGM absorption,
  assuming the galaxy lies at z=6.
 `Alt' indicates an alternate ID in \citet{2004MNRAS.355..374B}}
\label{tab:ewlims}
\end{table*}

\subsection{Implications of High Equivalent Widths} 
  \label{sec:EWimp}
  
  The presence of a tail of large equivalent width emission line sources has
  implications for the redshift distribution expected from an $i'$-drop 
  selection.
  
  The redshift distribution and number densities of $i'$-drop galaxies
  have usually been calculated assuming a negligible contribution from
  line emission to the measured magnitudes and colours. In the case of
  line emitters with rest frame equivalent widths $W_0<30$\AA\ this is
  a reasonable assumption, with the contribution to $z'$ flux
  amounting to a few percent.
  
\begin{figure}
\begin{center}
\resizebox{0.85\columnwidth}{!}{\includegraphics{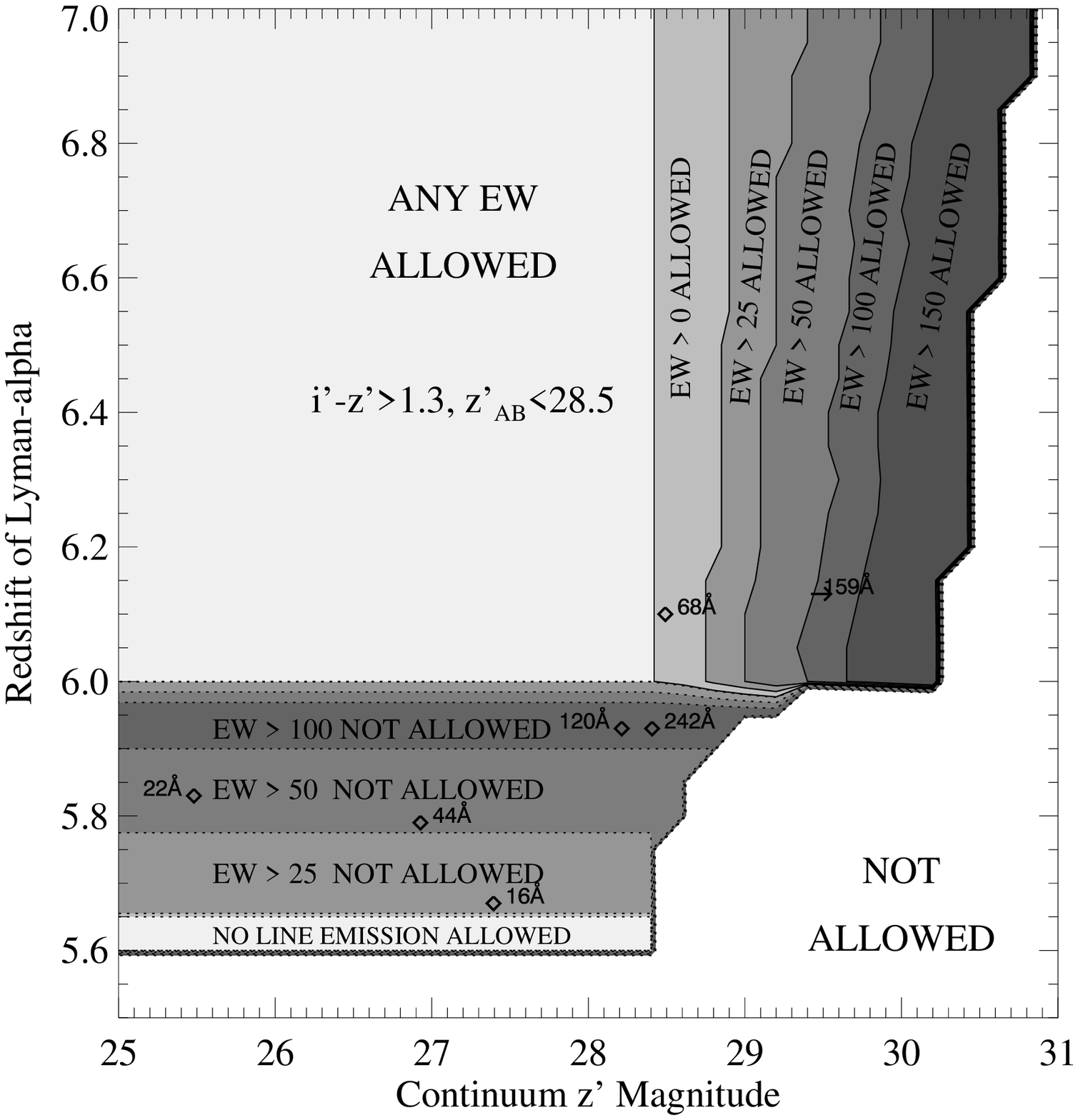}}
\end{center}
  \caption{The effects of Lyman-$\alpha$ equivalent width on the 
    $i$-drop selection function. Contours show the equivalent width
    line required to place a galaxy with the given redshift and
    continuum magnitude into the selection window. Solid lines
    indicate minimum equivalent width required, dotted lines indicate
    maximum equivalent width permitted. In the absence of line
    emission, all galaxies above the limiting observed magnitude and
    at $z>5.6$ are detected. In the presence of strong line emission,
    galaxies at $z<6.0$ fall out of the selection as the emission line
    lies in the $i'$-band, while fainter galaxies at high redshift can
    be promoted into the selection due to brightening of the measured
    $z'$ magnitude. The properties of the seven $i'$-drop sources with
    redshifts in this range are shown with diamonds.}\label{fig:ewsel}
% Notice to self: This figure comes from data2/GLARE/model_2006
\end{figure}

  For line emitters with larger equivalent widths there can be a
  significant effect on the selection function of $i'$-drop galaxies.
  This effect falls into three main regimes as illustrated by figure
  \ref{fig:ewsel}. If the emission line falls in the $i'$-band filter
  (i.e.  $z<6.0$), then the $i'-z'$ colour of the galaxy is reduced by
  the line emission, and sources with high equivalent widths fall out
  of the colour-magnitude selection window. A source at our $z'$
  detection limit with a rest frame Lyman-$\alpha$ equivalent width of
  100\AA can be as blue as $i'-z'=0.67$ (AB) and an intrinsic line
  width of 150\AA would lead to a colour of just $i'-z'=0.46$ (AB).
  These colours are similar to those of much lower redshift galaxies.
  Ensuring a complete selection of line emitting galaxies at
  $5.6<z<6.0$ is therefore impossible using a simple colour-selected
  sample without also including a great many lower redshift
  contaminant sources.

  If the line emission falls
  in $z'$ filter, the $i'-z'$ colour is enhanced and galaxies with
  continuum flux below the $z'$ limit are promoted into the selection
  window. Hence at $z>6.5$ the $i'$-drop criterion can select a
  population of low continuum, strong line emission sources rather
  than the continuum break sources it targets. There is also a
  redshift range in which the line emission would fall in the overlap
  region between filters and both effects compete.

This effect may lead to a bimodal redshift distribution for $i'$-drop
galaxies, with weak emission line sources preferentially selected
towards lower redshift, and high equivalent width sources selected at
higher redshifts. It is necessary to quantify the equivalent width
distribution, combining galaxies at the faint limits explored here
with brighter galaxies, in order to tightly constrain the galaxy
luminosity function at any given redshift.

As figure \ref{fig:ewsel} illustrates, our good line emission
candidates all lie in regions of parameter space affected by
equivalent width selection biases. Our candidates lie in parameter
space theoretically allowed for their equivalent width with two
exceptions - GLARE 1054 and 3011 both have strong emission lines in
the tail of the $i'$ filter that should have forced their $i'-z'$
colours out of our selection function. We note, however, that that
both of these sources have $i'-z'$ colours within 1\,$\sigma$ of our
minimum selection cutoff, and may have scattered up into out
selection. Only one source is boosted into the selection function by
virtue of its line emission rather than continuum flux, suggesting
that our sample is, as expected, dominated by continuum-selected
sources. If we are indeed missing sources with high emission line
equivalent widths at the low redshift end of our survey (where our
redshift selection function is at its peak), then our conclusion that
the distribution of line emission at high redshift extends to larger
equivalent widths is strengthened.

This effect has interesting implications for the luminosity functions
presented in the literature for $z\approx6$ galaxies. If the $i'$-drop
criteria are overestimating the number density of faint continuum
sources (due to contamination by a tail in the distribution of strong
line emitters) then it is possible that the faint end slope of the
luminosity function is shallower than hitherto reported.
 If there
is a large number of strong emitters at $5.6<z<6.0$ that have fallen
out of the colour selection, then the number density of sources at
these redshifts could be underreported. This may contribute towards
the discrepancy between the observed evolution in the LBG luminosity
function between $z=4$ and $z\approx6$
\citep[e.g.][]{2004MNRAS.355..374B,2004astro.ph..6562B} and the reported lack of evolution
in the Lyman-$\alpha$ emitter luminosity function over the same
redshift range \citep[e.g.][]{2004astro.ph..7408M}.

\begin{figure}
\begin{center}
\resizebox{0.9\columnwidth}{!}{\includegraphics{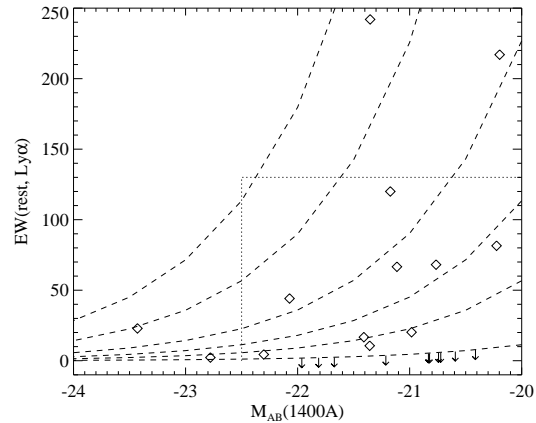}}
\end{center}
\caption{The equivalent width distribution of GLARE line emitters as a 
 function of absolute magnitude. The region of parameter space surveyed by
 the compilation of Ando et al (2006) is bordered by a dotted line. Dashed
 lines indicate the loci $z=6$ emission lines with luminosities of $10^{44}$,
  $5\times10^{43}$, $2\times10^{43}$, $10^{43}$, $5\times10^{42}$ and $10^{42}$
 ergs\,s$^{-1}$ (from top left to bottom right). Our equivalent width limit
 for sources without line detection corresponds to a rest frame luminosity of 
 $10^{42}$ ergs\,s$^{-1}$, assuming those sources lie at $z\approx6$.}
\label{fig:absmag}
\end{figure}

\citet{2006astro.ph..5289A} studied the equivalent width distribution of a
heterogeneous sample of $i'$-drop and narrowband selected galaxies with
spectroscopically confirmed Lyman-$\alpha$ emission at
$z\approx6$. They commented on an apparent dearth of UV-luminous
galaxies with strong Lyman-$\alpha$ emission lines, noting an increase
in the fraction of strong line emitters at $M_{AB}=-21$. In figure
\ref{fig:absmag} we show the equivalent distribution of the GLARE line
emitters as a function of absolute magnitude.

We detect five bright sources with a line luminosity brighter than
$2\times10^{43}$\,ergs\,s$^{-1}$, thus occupying a regime unoccupied
 by the Ando et al sample. Of these, two possible emission line
sources have equivalent widths greater than any in the previous sample.
The four sources with a calculated line luminosity between $5\times10^{43}$
and $10^{44}$\,ergs\,s$^{-1}$ are uniformly distributed in absolute magnitude.

Although the number statistics of our sample is small, our sample is
uniformly selected by broadband magnitude, removing possible biases
due to the combination of two methods. Narrow-band surveys are biased
towards sources with faint continuum and bright Lyman-$\alpha$
emission lines, whereas the $i'$-drop selection method is more
uniformly sensitive to Lyman-$\alpha$ emission except at the very
faint and low redshift ends of the sample. Our results do not support
those of Ando et al., although clearly further observations are
required to clarify the fraction of line emitters at bright
magnitudes. It is possible that the fraction of sources with high
equivalent width line emission is subject to environmental effects
(e.g. the local number density of galaxies). The small HUDF field is,
of course, subject to the effects of cosmic
variance\,\citep{2004MNRAS.355..374B}. For sources as clustered as
Lyman break galaxies at moderate redshifts are known to be, a variance
in number density of 40\% is expected in a field this size \citep{somerville04}.

\section{Discussion}
\label{sec:disc}

  \subsection{The Redshift Distribution of $i'$-drop Galaxies} 
  \label{sec:zdist}

The redshift distribution expected for line emitters detected by the
GLARE survey is the convolution of the selection functions imposed by
the $i'$-drop selection criteria and by the spectral response of
Gemini/GMOS in this configuration, modified by the probability of
detecting a line that falls on a night sky emission line.

\begin{figure}
\begin{center}
\resizebox{0.85\columnwidth}{!}{\includegraphics{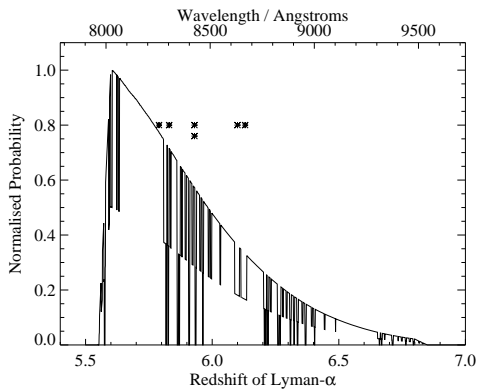}}
\end{center}
  \caption{
    The redshift distribution expected for sources with line detection
    in the GLARE spectroscopy. Assumed a distribution of magnitudes
    given by a Schecter luminosity function, with constant number
    density over the redshift interval $5.5<z<7.0$, modified by the
    sensitivity of Gemini/GMOS in our configuration, and assuming that
    a line is undetected if under a skyline, and has a 50\%
    probability of being detected if under a weaker skyline. The
    redshifts of our candidate line emitters are shown by asterisks,
    see section \ref{sec:zdist} for details.  }\label{fig:zdist}
\end{figure}

In figure \ref{fig:zdist} we show the expected distribution of
emission line redshifts, neglecting the influence of
strong line emission on source magnitude and colour (see above), and
assuming a Schecter (1976) luminosity function with parameters
$L^\ast=L^\ast$($z=3$), $\alpha=-1.9$ and
$\phi^\ast=\phi^\ast$($z=3$) (Stanway 2004). Changing the parameters of the
luminosity function has a slight effect on the shape of the
distribution, with a reduction in $L^\ast$ with increasing redshift
slightly broadening the peak of the distribution in redshift space.

We take the conservative assumption that a source has a detection
probability of 0\% if lying
on a very strong line ($>5\times$ the inter-line sky continuum)
and 50\% on top of a weaker sky line ($>1.5\times$ sky continuum
level). The night sky spectrum is measured directly from
our spectroscopy.

Clearly, our survey is most sensitive to sources lying towards the
lower-redshift end of our redshift range, and to sources lying in the
low OH line regions commonly exploited by Lyman-$\alpha$
surveys at $z\approx5.7$ and $z\approx6.6$. Nonetheless, we have a
reasonable probability of detecting sources in the skyline complexes
at $z<6.5$, particularly if the source emission lies between skylines.

Given this theoretical sensitivity distribution, it is interesting to
note that the five robust line emitters presented in this paper (and
one possible emission line also lying in this redshift range) do not
fall at the peak of the detection sensitivity distribution, but rather
within the skyline complexes. In \citet{2004ApJ...604L..13S} we
suggested weak evidence for large scale structure in the HUDF field at
$z=5.8$, based on the three redshifts known at the time. More recently
\citet{2004astro.ph..3458P} have found evidence for an over-density at
$z=5.9\pm0.2$ in the same field based on GRAPES Grism spectroscopy.
Given that it unlikely that our fainter targets will yield redshifts
with the current generation of spectrograph, confirmation of this may
prove challenging. However, our results so far appear consistent with
large scale structure at $z\approx5.8$ in this field.

\subsection{The Space Density of Lyman-$\alpha$ Emitters}
\label{sec:LAEs}
   
We have identified a total of 12 (5 of which were strong) candidate
emission lines with a luminosity in the range
$(1-50)\times10^{42}$\,ergs\,s$^{-1}$ in our sample of 24 $i'$-drop
galaxy candidates. If this is indicative of the $i'$-drop selected
galaxy population as a whole, and if the equivalent width distribution
is invariant with magnitude, we would expect to observe line emission
of this strength in approximately 25 (11 strong) of the 51 $i'$-drop
galaxies identified in the HUDF by \citet{2004MNRAS.355..374B} (which
were used by GLARE for candidate selection). Hence, this suggests a
line emitter space density of $6.7 (2.9) \times10^{-4}$\,($\Delta
log{L}$)$^{-1}$\,Mpc$^{-3}$), using the effective comoving volume for
the $i'$-drop selection in the HUDF survey of
$2.6\times10^{4}$\,Mpc$^{3}$. This space density comparable to the
space density of $6.3\times10^{-4}$\,($\Delta
log{L}$)$^{-1}$\,Mpc$^{-3}$ estimated by \citet{2006PASJ...58..313S}
for Lyman-alpha emitters of comparable luminosity
($(3-25)\times10^{42}$) at $z\approx5.7$ in the Subaru Deep Field.

%, even when only the most
%confident line identifications are used for the inference.
 
Similarly, at $z=5.7$ the Lyman-Alpha Large Area Survey (LALA) found a %typical
number density of $3.2\times10^{-4}$\,Mpc$^{-3}$ for Lyman-$\alpha$ emitters
with $L>10^{42}$\,erg\,s$^{-1}$ \citep{2004ApJ...617L...5M}.  This
would predict the detection of 7 Lyman-$\alpha$ emitters to our
detection limit in our survey volume. This is consistent with, although
slightly lower, than the number of emitters we detect.

Inferring the space density of emitters from the relatively small
sample in this survey is, of course, subject to large errors due to
Poisson noise on the small number statistics.  However, at least two
physical explanations are possible for a discrepency between the line
emitter density found by GLARE and by LAE surveys. As mentioned in section
\ref{sec:zdist}, the HUDF is in a region with a known overdensity of
sources at $5.7<z<5.9$. GLARE is also likely to detect a higher
apparent density of line emitters if the Lyman-$\alpha$ photon escape
fraction is higher in the relatively faint galaxies studied here than
in more massive galaxies. If this is true then the number of sources
observed at faint magnitudes exceeds the number predicted from
brighter surveys. This phenomenon may well be expected if the deeper
potential well of more massive galaxies traps more dust and a denser
ISM than that existing in the shallow potential wells of small, faint
galaxies.

  \subsection{The Emission line properties of sub-L$^\ast$ Galaxies} 
  \label{sec:sub_lstar}
  
  This spectroscopy is the longest integrated exposure on a single
  mask using an 8m telescope. It reaches the faintest line limits
  available for a large sample with a homogeneous $i'$-drop colour
  cut.  The 3 sigma limit in most slits is approx
  $4.1\times10^{-19}$\,erg\,cm$^{-2}$\,s$^{-1}$\,\AA$^{-1}$. 
%However,
%  the requirement that images are detected independently in both nod
%  positions. 
  
  Hence we are able to observe line emissions from candidates
  significantly fainter than other surveys targeting $i'$-drop
  galaxies.
  
  Given the expenditure in telescope resources required to reach these
  faint limits, it is likely that many of the HUDF candidates may well
  remain inaccessible to ground based spectroscopy until the advent of
  the Extremely Large Telescopes, and from space by JWST with NIRSpec.
  HUDF $i'$-drops can be identified with reasonable confidence to
  $z'_{AB}<28.5$ (Bunker et al.\ 2004). A source with $z'_{AB}=28.5$
  and an emission line of W=20\AA\ (rest) at $z=6$, with complete
  Lyman-$\alpha$ absorption shortward of the line, would have a
  contamination fraction of 13.5\% to the $z'$-band from this line,
  and a continuum flux of
  $6.4\times10^{-21}$\,erg\,cm$^{-2}$\,s$^{-1}$\,\AA$^{-1}$ -- i.e. a
  continuum flux a factor of $\sim 10$ less than we are able to detect
  at the $1\,\sigma$ level per spectral resolution element (6.5\,\AA )
  .  However, such a source would possess a line flux of
  $6\times10^{-18}$\,erg\,cm$^{-2}$\,s$^{-1}$, well within our
  detection limits for emission lines.
  
  Hence, although emission lines are detectable in surveys such as
  this, even for the faintest of our target sources, continuum spectra
  will remain inaccessible for some time to come. Even GLARE 1042, the
  brightest $z\approx6$ in the HUDF does not have sufficient signal to
  noise to in this spectroscopy to detect absorption features, even
  though the continuum is now detected at high significance. The
  detection of stellar absorption lines would provide valuable
  kinematic information on the galaxies and their outflows, however
  while such investigations may be possible on a stacked image of our
  faint spectra, it is impossible on individual slits. A stack of the
  faint GLARE spectra will be considered in a future paper.

 % \textit{Look at stack? Are the redshifts of the faint sources secure
 %   enough? Can we say anything about velocity structure of the LAE
 %   line? Check the instrumental resolution}

\section{Conclusions}
\label{sec:conc}

In this paper we have presented spectroscopy of the faintest known
sample of $i'$-drop galaxies, derived from the Hubble Ultra Deep
Field. An exposure of 36 hours per target was obtained with the
Gemini-South telescope.

Our main conclusions can be summarised as follows:

i) We have obtained extremely deep spectroscopy for twenty-nine
science targets, reaching $3\,\sigma$ line flux limits of
$2.5\times10^{-18}$\,erg\,cm$^{-2}$\,s$^{-1}$ -- corresponding to
equivalent width limits of $<6$\,\AA\ for the majority of our targets.

ii) We identify five $i'$-drop galaxies as good Lyman-$\alpha$
emitter candidates, four as possible candidates, and
a further five tentative emission lines which we consider unlikely to be real.

iii) We have considered the observed equivalent width distribution of
$i'$-drop galaxies in the HUDF, and identify a tail of line emitters
with very high equivalent widths which is not seen in the lower-redshift
Lyman-break galaxies at $z\approx 3-4$. 

iv) Several possible explanations for this effect exist. These include
a tendency towards stronger line emission in faint sources, extreme
youth or low metallicity in the Lyman-break population at high
redshift, or possibly a top-heavy initial mass function.

v) At the low redshift end of our selection function ($5.6<z<6.0$),
the $i'$-drop selection method will fail to select line emitters with
high equivalent width due to line contamination producing blue
colours.  In contrast, at the high redshifts end ($z>6$),
continuum-faint line emitters may enter the selection function if the
Lyman-$\alpha$ line is sufficiently luminous.  This has implications
for the redshift and continuum magnitude selection function of
$i'$-drop galaxies.

vi) This sample significantly increases the number of faint $i'$-drop
galaxies with known redshifts, and may begin to bridge the divide
between continuum and line selected galaxies at $z\approx6$.

\section*{Acknowledgments}

Based on observations obtained at the Gemini observatory, which is
operated by AURA Inc, under a cooperative agreement with the NSF, on
behalf of the Gemini partnership. Also based in part on observations
with the NASA/ESA Hubble Space Telescope, obtained at the Space
Telescope Science Institute which is operated by AURA Inc under NASA
contract NAS 5-26555. We thank the GOODS and HUDF teams for promptly
releasing their valuable surveys to the community. We also thank all
members of the GLARE collaboration.

ERS gratefully acknowledges support from NSF grant AST 02-39425. AJB
acknowledges support from a Philip Leverhulme Prize.

We thank the anonymous reviewer for their helpful comments which have enhanced this paper.

\label{lastpage}

%\bibliographystyle{mn2e}
%\bibliography{/usr/users/stanway/data/thesis/refs/ref}

\end{document}